\documentclass[twocolumn,superscriptaddress,amsmath,amssymb,floatfix,aps,notitlepage,nokeys,pre,longbibliography]{revtex4-1}

\usepackage{mathrsfs, hyperref}
\usepackage{amssymb, amsbsy, amsmath, latexsym, dsfont, array, layout, graphics,mathrsfs,braket,amsfonts,amsthm,
  amssymb,graphicx,subfigure, youngtab,color,bm,mathtools,braket,verbatim,url,cleveref,natbib,hypernat,extarrows,inputenc,multirow}
\usepackage[usernames,dvipsnames]{xcolor}

\usepackage{graphicx}
\usepackage{dcolumn}
\usepackage{bm}
\usepackage{amsmath}
\usepackage{natbib}
\DeclareMathOperator{\arcsinh}{arcsinh}

\begin{document}


\title{Mechanical Properties Of Fluctuating Elastic Membranes Under Uni-Axial Tension}
\author{Mohamed El Hedi Bahri}
\email{mbahri@princeton.edu}
\affiliation{Department of Mechanical and Aerospace Engineering, Princeton University, Princeton, NJ 08544, USA}
\author{Siddhartha Sarkar}
\email{sarkarsi@umich.edu}
\affiliation{Department of Physics, University of Michigan, Ann Arbor, MI 48103, USA}
\author{Andrej Ko\v{s}mrlj}
\email{andrej@princeton.edu}
\affiliation{Department of Mechanical and Aerospace Engineering, Princeton University, Princeton, NJ 08544, USA}
\affiliation{Princeton Institute for the Science and Technology of Materials, Princeton University, Princeton, NJ 08544, USA}

\date{\today}

\begin{abstract}
Atomically thin sheets, such as graphene, are widely used in nanotechnology. Recently they have also been used in applications including kirigami and self-folding origami, where it becomes important to understand how they respond to external loads. Motivated by this, we investigate how isotropic sheets respond to uniaxial tension by employing the self-consistent screening analysis method and molecular dynamics simulations. Previously, it was shown that for freely suspended sheets thermal fluctuations effectively renormalize elastic constants, which become scale-dependent beyond a characteristic thermal length scale (a few nanometers for graphene at room temperature), beyond which the bending rigidity increases, while the in-plane elastic constants reduce with universal power law exponents. For sheets under uniaxial tension, $\sigma_{11}$, we find that beyond a stress-dependent length scale, the effective in-plane elastic constants become strongly anisotropic and scale differently along the axis of uni-axial stress and orthogonal to it. The bending rigidities on the other hand will not exhibit any anomalous behavior beyond this stress-dependent length scale. In addition, for moderate tensions we find a universal non-linear stress-strain relation. For large uni-axial tensions, the Young's modulus of the bare elastic material is recovered.
\end{abstract}

\maketitle


\section{\label{sec:level1}Introduction}

Geometric non-linearities are the mainstay of the study of the mechanics of slender structures \cite{timoshenko1959theory}. Though this field is quite old, only over the last few decades have the effects of temperature on the mechanics of 2-D materials been studied, as is exemplified by polymerized membranes, graphene and a whole host of other 2D materials such as $\text{BN}$, $\text{WS}_2$ and $\text{MoS}_2$ which have been discovered over the last decade \cite{kantor1987crumpling,nelson1987fluctuations, aronovitz1988fluctuations, guitter1989thermodynamical,novoselov2005two,katsnelson2007graphene, amorim2014thermodynamics}. Free-standing layers of these 2D crystals offer an experimentally realizable system for exploring how mechanical behavior of thermalized elastic membranes. Further manipulation of these 2D crystals for the creation of metamaterials generates new opportunities for research on the interface of mechanical and electronic properties of 2D crystals. One such recent example shows the experimental realization of kirigami graphene where large effective strains did not affect its conductivity \cite{blees2015graphene}.

Although, 2-D elastic crystals may be viewed as a higher dimensional extension of the $D=1$ elastic polymer, there are some major differences between these two physical systems. Analogous to the persistence length  of thermalized polymers (the length scale over which a polymer is approximately straight) \cite{de1979scaling}, 2-D materials have a temperature-dependent length scale, named the thermal length scale, $\ell_{\text{th}}$, beyond which temperature plays a role in elastic responses to external stresses. However, due to the coupling between the Goldstone flexural phonons and the in-plane phonons, 2D elastic materials of arbitrarily large size avoid being subject to the Mermin-Wagner theorem when $D=2$ \cite{nelson1987fluctuations, coquand2019spontaneous,paczuski1988landau} and thus remain flat at sufficiently low temperatures even beyond this thermal length scale. The result is a mean-field flat phase below a crumpling temperature that gives rise to elastic moduli that exhibit anomalous scale dependence. In \cite{nelson1987fluctuations, aronovitz1988fluctuations,guitter1989thermodynamical} it was shown that beyond the thermal length scale, the effective Lam\'e constants scaled as $\lambda_R(q),\mu_R(q) \sim q^{\eta_u}$ (where $q$ is the Fourier scale and $\eta_u \approx .4$). On the other hand the effective bending rigidity divered as $\kappa_R(q) \sim q^{-\eta}$ ($\eta \approx .8$). These results for the numerical exponents have been recently verified to 2 and 3-loop order \cite{metayer2022three,coquand2020flat,mauri2020scaling,mermin1966absence,mermin1968crystalline,halperin2019hohenberg}. In addition, the arguments demonstrating that elasticity in $D$-dimensions exhibits scale but not conformal invariance \cite{riva2005scale} was extended to the case of $D$-dimensional elastic membranes embedded in $D+d_c$ dimensions (where $d_c$ is the co-dimension) \cite{mauri2021scale}.

Experimental measurements of the scale-dependence of the elastic moduli of thermalized membranes and the resulting mechanical properties (such as the non-linear relation $\epsilon \sim \sigma^{\eta/(2-\eta)} $) in the absence of quenched disorder have not been realized yet.  However, many theoretical and simulation efforts have been realized to extend the original results of \cite{nelson1987fluctuations} to a variety of interesting cases and to understand the mechanical response of 2-D materials. In particular, for isotropically stressed fluctuating membranes, when stresses are larger than the linear response but less than one that would flatten out all thermal wrinkles, a non-linear relation between stress and strain was obtained $\epsilon \sim \sigma^{\eta/(2-\eta)}$ \cite{kovsmrlj2016response, aronovitz1989fluctuations}. Thermal fluctuations also increase critical buckling load with respect to the Euler buckling load due to the divergent effective bending rigidity $\kappa_R(q) \sim q^{-\eta}$ \cite{morshedifard2021buckling,hanakata2021thermal}. Extension of the theory to inversion-asymmetric tethered membranes (such as graphene coated with a material on one side) has been recently done in which a double spiral phase and long range orientational order was predicted \cite{banerjee2019rolled}. Realistic considerations of single clamped boundary conditions have also been reported to introduce a spontaneous symmetry-breaking tilt \cite{chen2022spontaneous}. A recent extension of the theory of mono-layer elastic membranes to bi-layers was also done and found that the effective scaling of the elastic moduli did not change in the infrared limit ($q \rightarrow 0$) \cite{mauri2020scaling}. In addition, a new universality class was obtained with different anomalous elastic exponents have been done in the presence of an external field that breaks the rotational symmetry of the embedding space \cite{le2021thermal}. Although early theoretical studies focusing on estimating the Poisson ratio of stress-free membranes found a universal value of $-1/3$ \cite{le1992self,falcioni1997poisson}, other more recent studies \cite{burmistrov2018stress,saykin2020absolute,los2016scaling} indicate that this may not be the case. Thus more investigations will be necessary to further comprehend the Poisson ratio as a function of exerted stress. Further simulations have been done in an effort to consider the effect of experimental realities such as the quenched rippling of graphene and defects \cite{qin2017negative}. These simulations showed that the Poisson ratio decreased with aspect ratio between the amplitude of the ripples and the system size, even making it negative.

Other investigations have been conducted for elastic membranes with an intrinsic anisotropy \cite{radzihovsky1995new, radzihovsky1998elasticity}. For sufficiently high temperatures, due to the anisotropy, the mean field flat phase becomes un-stable and leads to a mean-field tubule phase, neglecting self-avoidance. This was further confirmed via non-perturbative approaches \cite{essafi2011crumpled}, which also better characterized the critical exponents associated with the phase transition. Anisotropies can be quite generic and thus the authors chose to focus on a tubule with effectively straight along one axis and crumpled along the other $D-1$ axes \cite{radzihovsky1995new}. Within the tubule phase, assuming no self-avoidance, the effective elastic moduli become scale-dependent but with a behavior that is different from that found in the flat phase. Simulations done in \cite{bowick1997numerical} confirmed the existence of this flat-to-tubule phase transition by inserting an anisotropy in the bending rigidities. They further measured the gyration radius as a function of the length of the tubule and obtained the scaling $R_G \sim L^{\nu_F}$ where the Flory exponent is $\nu_F \approx .3$, and found it to be within close agreement with the theory, $\nu_F=1/4$. 

\begin{figure}
\includegraphics[width=.48\textwidth]{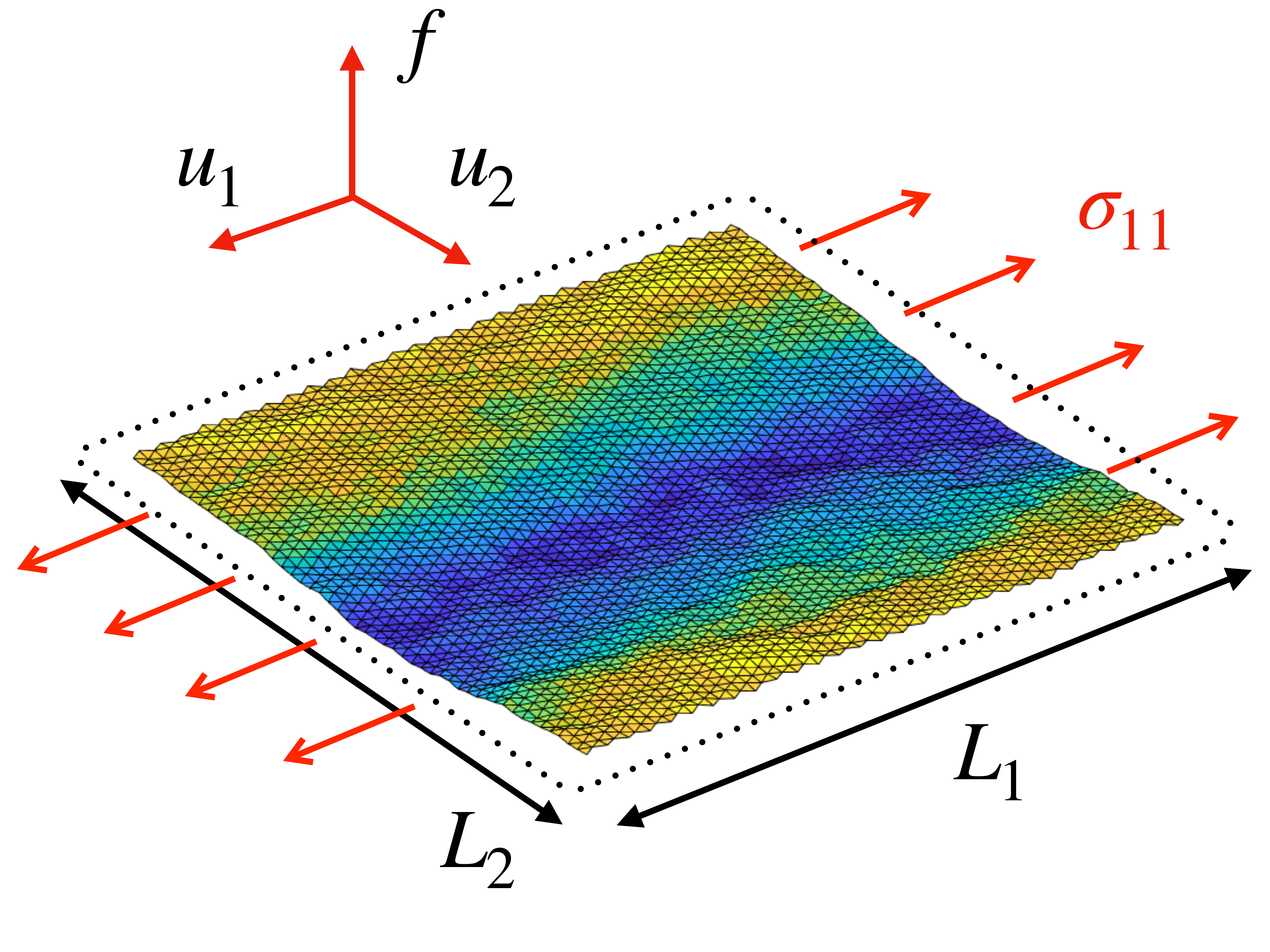}
\caption{A snapshot of a simulation of a thermally fluctuating sheet placed under a uni-axial tension. $u_i$ indicate the in-plane displacements whereas $f$ is the flexural/out-of-plane field.  The coloring of the membrane shows the scalar value of the height field (high frequency colors such as blue showing negative heights and low frequency colors such as yellow showing positive heights with respect to a zero-mean height). When a uni-axial stress is significant, transverse flexural fluctuations dominate as can be seen from the height coloration of the figure taken from a simulation. The dotted line marks the fact the $T=0$ size which shows that elastic sheets shrink when temperature is present.}
\label{fig:schematic}
\end{figure}

In this paper we focus on extending the theory of thermalized 2D elastic membranes to a scenario of physical interest in which a homogeneous uni-axial tension is exerted. A snapshot from a simulation can be seen in Fig.~\ref{fig:schematic} and illustrates the physical scenario. Stress will introduce a new wave-vector, $q_{\sigma}$, which will render the theory dependent on the its relative magnitude with respect to $q_{\text{th}} = 2 \pi / \ell_{\text{th}}$. We will explore the scaling of these elastic moduli at a variety of length scales and show an anomalous scaling at high stresses and temperatures that becomes identical to that of \cite{radzihovsky1995new, radzihovsky1998elasticity} in the tubule phase. In the infrared limit ($q\rightarrow 0$) the modulus $C_{2222}^R(q) \sim q$ whereas $C_{1111}^R \sim \text{constant}$, thus the system will exhibit strong anisotropy in the in-plane correlation functions. In the same limit, the moduli characterizing bending rigidities will exhibit anomalous behavior. Furthermore, as in the case of isotropic stress, we once again obtain a regime with a non-linear stress-strain relation, $\epsilon \sim \sigma^{\eta/(2-\eta)}$, when $2\pi/L<q_{\sigma}<q_{\text{th}}$ (where $L$ is the system size).

We will first begin in Sec.~\ref{sec:level2} by introducing the theory in the absence of stress and explore the consequences of introducing uni-axial stresses for the symmetries of the free energy as well as the appearance of a new length scale beyond which stress becomes important. In Sec.~\ref{sec:ScaleBehavior}, we study uni-axially stressed membranes via an engineering dimension analysis and Self-Consistent-Screening-Analysis (SCSA) equations to obtain the scaling of effective elastic moduli. Simulations performed in the NPT ensemble using the LAMMPS package confirm the scaling of this theory. In Sec.~\ref{sec:StressStrain}, we will study stress-strain relations using what we will have learned about the elastic moduli.

\section{\label{sec:level2}Statistical Mechanics Of Elastic Membranes}
We first discuss the free energy function of a general $D$-dimensional elastic membrane embedded in ($D$+1)-dimensions undergoing small deformations with respect to the reference flat state. Using Einstein notation in which repeated indices are implicitly summed over, such a function has the form~\cite{landau1995course}:
\begin{equation}
\label{eq:FreeEnergy}
\begin{split}
    \mathcal{F} = \frac{1}{2} \int d^D\mathbf{r} [\lambda u_{ii}^2 + 2 \mu u_{ij}^2 +\kappa K_{ii}^2 -2\kappa_G \text{det}(K_{ij})] 
\end{split}
\end{equation}
where $\lambda$ and $\mu$ are the elastic Lam\'e constants and $\kappa$ and $\kappa_G$ are the bending and Gaussian bending rigidities. Here we use the strain and curvature tensor equations:
\begin{equation}
    \begin{split}
        & u_{ij} = \frac{1}{2} [\partial_i u_j + \partial_j u_i+ \partial_i f \partial_j f ] \\
        & K_{ij} = \partial_i \partial_j f 
    \end{split}
\end{equation}
where the indices $i,j$ run through the $D$ intrinsic dimensions of the elastic membrane. These describe the deformations from a reference flat metric and zero-curvature state, with $u_i$ being the in-plane displacements along the i-th axis and $f$ being the out-of-plane displacement. The strain tensor $u_{ij}$ expresses stretching and shearing whereas $K_{ij}$ expresses curvatures. Note that we have omitted $(\partial u)^2$ from the non-harmonic portion of the strain tensor due to the fact that in-plane stretching costs more energy than stretching due to the out-of-plane deformations represented by $(\partial f)^2$. By means of an engineering dimension analysis done in Sec.~\ref{sec:ScaleBehavior}, one can show that $(\partial u)^2$ is irrelevant and can thus be ignored.

The effect of thermal fluctuations in a system with free energy function $\mathcal{F}$ can be extracted from the correlation functions, obtained via functional integrals over all membrane configurations \cite{kardar2007statistical}:
\begin{equation}
\begin{split}
     \mathcal{G}^R_{u_iu_j}(\mathbf{r}_2-\mathbf{r}_1) &= \frac{1}{\mathcal{Z}} \int \mathcal{D}[u_i,f] u_i(\mathbf{r}_2)u_j(\mathbf{r}_1) e^{-\mathcal{F}/k_BT} \\
         \mathcal{G}^R_{ff}(\mathbf{r}_2-\mathbf{r}_1) &= \frac{1}{\mathcal{Z}} \int \mathcal{D}[u_i,f] f(\mathbf{r}_2)f(\mathbf{r}_1) e^{-\mathcal{F}/k_BT}
        \end{split} 
\end{equation}
where $e^{-\mathcal{F}/k_BT}$ is the temperature dependent Boltzman weight and $\mathcal{Z}$ is the normalizing partition function, $\mathcal{Z}= \int \mathcal{D}[u_i,f]  e^{-\mathcal{F}/k_BT}$. Due to the form of the in-plane strain tensor, the free energy function is not harmonic in the displacement parameters $u_i$ and $f$. In the absence of such an-harmonic terms and stress and under periodic boundary conditions (so we may integrate out the Gaussian bending term via the Gauss-Bonnet theorem), the correlation function of the flexural phonons  and in-plane phonons of a system of Fourier scale $q$ take the form \cite{nelson1987fluctuations,nelson2004statistical}:
\begin{equation}
\label{eq:HarmonicInPlane}
    \mathcal{G}_{u_iu_j}(\mathbf{q}) = \frac{k_BT P^T_{ij}(\mathbf{q})}{A\mu q^2}+\frac{k_BT (\delta_{ij}-P^T_{ij}(\mathbf{q}))}{A(2\mu+\lambda)q^2}
\end{equation}
\begin{equation}
\label{eq:HarmonicFlexural}
    \mathcal{G}_{ff}(\mathbf{q}) = \frac{k_BT}{A \kappa q^4}
\end{equation}
where $A$ is the membrane area and $P^T_{ij}(q) = \delta_{ij} -q_i q_j/q^2$ is the transverse projection operator. We have taken the form of the Fourier transform to be: $G(\mathbf{r}) = G(0)+\sum_{\Lambda \geq |\mathbf{q}|\geq 2 \pi/L } G(\mathbf{q}) e^{i \mathbf{q} \cdot \mathbf{r}} = G(0)+\int_{2\pi/L}^{\Lambda} \frac{d^2\mathbf{q}}{A} G(\mathbf{q})e^{i \mathbf{q} \cdot \mathbf{r}} $. In this definition of the Fourier transform, $\Lambda$ is the UV cutoff introduced by the microscopic scale of the system where the continuum elastic theory breaks down i.e. $\Lambda = \pi/a$ where $a$ is the lattice spacing of the material. The correspondence between real lengths and Fourier space inverse lengths is taken to be: $q = 2 \pi / \ell$. Via scaling analysis, a length scale subtly emerges due to the presence of temperature. It is well known from plate theory that an-harmonic terms play a role once the magnitude of deflection becomes comparable to the thickness of the plate \cite{landau1995course}. Considering specifically $D=2$ materials such as graphene, though it is atomically thin, we can assign an effective thickness, derivable via the elastic formula, $t \sim \sqrt{\kappa/Y}$ where $Y$ is the 2D Young's modulus ($Y = 4 \mu (\lambda+\mu)/(\lambda+2\mu)$) \cite{landau1995course}. Assuming we take a plate thickness $t$ under-going an out-of-plane displacement deformation of amplitude $f$ over a length $L$, a scaling analysis can be performed to compare the bending energy and stretching energy cost from Eq.~\eqref{eq:FreeEnergy} \cite{landau1995course}:
\begin{equation}
\label{eq:ThermalLengthDer}
    \begin{split}
        &\mathcal{F}_{\text{stretching}} \sim Y \bigg( \frac{f^2}{L^2} \bigg)^2 \\ 
        &\mathcal{F}_{\text{bending}} \sim Y t^2 \bigg( \frac{f}{L^2} \bigg)^2
    \end{split}
\end{equation}
When the two energy costs are of comparable order, an-harmonic terms can no longer be ignored. Indeed, one can notice from the form of Eq.~\eqref{eq:ThermalLengthDer} that this occurs when $f \approx t \sim \sqrt{\kappa/Y}$. Inserting this equivalence of length scales and the Fourier form $q \sim 1/L$ into Eq.~\eqref{eq:HarmonicFlexural} and solving for $L$, we obtain a length scale $\ell_{\text{th}} \sim \sqrt{\frac{\kappa^2}{Yk_BT}}$ when $D=2$. In general $D$ dimensions $\ell_{\text{th}} \sim (\frac{\kappa^2( \lambda+2\mu)}{4 \mu ( \lambda+\mu)k_BT})^{\frac{1}{4-D}}$ \cite{radzihovsky1991statistical} which can be derived by using the general effective thickness $t \sim \sqrt{\kappa(\lambda+2\mu)/(4\mu(\lambda+\mu))}$. This defines the thermal length scale, beyond which temperature affects the mechanical properties of the elastic membrane and an-harmonic terms can no longer be ignored.

The non-linear form of the in-plane strain tensor produces long-ranged coupling of Gaussian curvatures and induces a non-trivial scaling of the correlation functions beyond the thermal length scale, $\ell_{\text{th}}$ \cite{coquand2019spontaneous,nelson1987fluctuations,nelson2004statistical}. In the absence of stress, the scaling of the correlation functions is known in the long-wavelength limit and temperature renormalizes the moduli and renders them scale-dependent \cite{aronovitz1988fluctuations, radzihovsky1991statistical, nelson1987fluctuations}:
\begin{widetext}
\begin{equation}
\label{eq:IsoGreensFunction}
    \begin{split}
        \mathcal{G}_{u_iu_j}^R(\mathbf{q})  &= \frac{k_BT P^T_{ij}(\mathbf{q})}{A\mu_R(q)q^2}+\frac{k_BT (\delta_{ij}-P^T_{ij}(\mathbf{q}))}{A(2\mu_R(q)+\lambda_R(q))q^2}  
        \sim \bigg( \frac{q}{q_{\text{th}}}\bigg)^{-2-\eta_u} \\
        \mathcal{G}_{ff}^R(\mathbf{q}) &= \frac{k_BT}{A \kappa_R(q) q^4} \sim \bigg( \frac{q}{q_{\text{th}}}\bigg)^{-4+\eta} 
    \end{split}
\end{equation}
\end{widetext}
where $q_{\text{th}} = 2 \pi/\ell_{\text{th}}$. The anomalous exponents take the approximate values $\eta \approx .8$ and $\eta_u \approx .4$ for $D=2$ \cite{kovsmrlj2016response}. These exponents are not distinct but are related due to the form of the in-plane strain tensor $u_{ij}$, where $\partial u$ and $(\partial f)^2$ must scale together \cite{guitter1989thermodynamical}. This leads to the exponent identity $2\eta +\eta_u=4-D$ \cite{kovsmrlj2016response,guitter1989thermodynamical}. It is important to note that $D_{\text{uc}}=4$ is the upper critical dimension of the theory and thus no anomalous exponents will be present when $D>D_{\text{uc}}$.  For $D<D_{\text{uc}}$, these exponents imply that the renormalized bending rigidity diverges as $\kappa_R(q) \sim (q/q_{\text{th}})^{-\eta}$ whereas the renormalized in-plane moduli converge to zero as $\mu_R(q),\lambda_R(q),Y_R(q) \sim (q/q_{\text{th}})^{\eta_u}$ \cite{nelson1987fluctuations}.

Naively, in the presence of an arbitrary edge stress, $\sigma_{ij}$, applied to an edge with unit normal $\hat{\mathbf{n}}$, one would write the following free energy function \cite{kovsmrlj2016response}:
\begin{equation}
\label{eq:FreeEnergyStress}
\begin{split}
    \mathcal{F} = &\frac{1}{2} \int d^D\mathbf{r} [\lambda u_{ii}^2 + 2 \mu u_{ij}^2 +\kappa K_{ii}^2 -2\kappa_G \text{det}(K_{ij})] \\&- \oint d^{D-1}\mathbf{S}\, \hat{n}_i \sigma_{ij} u_{j}
\end{split}
\end{equation}
where the boundary term expresses the work done by an external stress. However, the effective theory at a given scale $\ell^* = 2\pi/q^*$ can be extracted by integrating out faster small-scale fluctuations. This can be done by splitting the phononic fields into pieces: $g_<(\mathbf{r}) = \sum_{|\mathbf{q}|<q^*} e^{i \mathbf{q} \cdot \mathbf{r}} g(\mathbf{q})$ and $g_>(\mathbf{r}) = \sum_{\Lambda>|\mathbf{q}|>q^*} e^{i \mathbf{q} \cdot \mathbf{r}} g(\mathbf{q})$ where $g \in \{u_i,f \}$ and integrating out the latter, $g_>$. By performing this integration we obtain \cite{kardar2007statistical}:
\begin{equation}
\label{eq:CoarseGrain}
\begin{split}
    & \mathcal{F}_{\ell^*}[u_{i<},f_<] = \\ & \quad \quad -k_BT \,\text{ln} \int \mathcal{D}[u_{i>},f_>] e^{-\mathcal{F}_a[u_{i<},f_<,u_{i>},f_>]/k_BT}
\end{split}
\end{equation}
where $\mathcal{F}_a[u_{i<},f_<,u_{i>},f_>]$ is the full free energy function without any phononic modes having been integrated out. Trivially, homogeneous isotropic stress will not cause renormalized anisotropies to develop since the stress will not break any rotational or mirror symmetries in the free energy. However, in the presence of a homogeneous anisotropic stress the isotropy of the free energy will be broken.


Thus it must be considered that these moduli can develop effective anisotropies for a non-isotropic stress and not only a scale dependence due to temperature. Despite the need for a generalization of the free energy, some symmetries will remain assuming the form of the stress to be ``uni-axial''.  We clarify here that in general $D$-dimensions we define ``uni-axial'' stress as the case in which all axes experience an equal tension $\sigma_{\alpha \alpha}=\sigma$ with $\alpha \in \{ 1, ...,D-1\}$ and $\sigma_{DD}=0$ (we call this the case of ``uni-axial'' tension since in $D=2$, which is the case of interest, it is indeed the correct physical scenario). As a brief aside, note here that we will use Greek letters to range over indices between $\{1,...,D-1\}$ and Roman letters as indices that range over $\{1,...,D\}$. We take this unusual definition of ``uni-axial'' to mimic the exact same theoretical formulation of tubules in Radzihovsky and Toner's theory \cite{radzihovsky1995new, radzihovsky1998elasticity}. Examining Eq.~\eqref{eq:FreeEnergyStress}, one can see that for ``uniaxial'' stresses the free energy will have at least orthorhombic symmetry; those being $D$ mirror symmetries across each of the $D$ axes. These orthorhombic symmetries will remain in renormalized free energies.

Accommodating orthorhombic anisotropy into the free energy, the generalization takes the form \cite{landau1995course}:
\begin{equation}
\label{eq:FreeEnergyAniso}
\begin{split}
    \mathcal{F} & =  \frac{1}{2} \int d^D\mathbf{r} [C_{ijkl} u_{ij} u_{kl} +B_{ijkl}K_{ij}K_{kl}] \\ & - \oint d^{D-1} \mathbf{S} \hat{n}_i \sigma_{ij} u_{j}
\end{split}
\end{equation}
where, the bare elastic moduli tensors have the fundamental major and minor symmetries: $C_{ijkl} = C_{klij}=C_{jikl}=C_{ijlk}$ and $B_{ijkl} = B_{klij}=B_{jikl}=B_{ijlk}$ \cite{ashcroft2022solid}. In addition to these, the orthorhombic symmetries will enforce that $C_{iiij} = C_{iijk} = C_{ijkl} = 0$ where each distinct index is taken to be a distinct number between $1$ and $D$. The same will hold true for the $B_{ijkl}$ tensor. In this notation, an isotropic elastic material will have the following decomposition: $C_{ijkl}=\lambda \delta_{ij} \delta_{kl} + \mu [\delta_{ik} \delta_{jl} +  \delta_{il} \delta_{jk}]$ and $B_{ijkl} = (\kappa-\kappa_G) \delta_{ij}\delta_{kl}+\kappa_G/2 [\delta_{ik}\delta_{jl}+\delta_{il}\delta_{jk}]$.

Under stress the effective flexural phonon correlation function may be defined:
\begin{equation}
\label{eq:EffectiveGreenFlexural}
    \mathcal{G}_{ff}^R(\mathbf{q}) = \frac{k_BT}{A[B_{ijkl}^R(\mathbf{q})q_iq_jq_kq_l +\sigma_{ij}q_iq_j]}
\end{equation}
whereas the correlation function for in-plane phonons takes the form:
\begin{equation}
\label{eq:GreenInPlanePhonon}
\mathcal{G}_{u_i u_j}^R = \frac{k_BT}{A} [C_{ikjl}^R(\mathbf{q})q_kq_l]^{-1}
\end{equation}
These renormalized elastic constants have been defined based on the analogous correlation functions in the harmonic approximation similarly as one would define $\lambda^R(q),\mu^R(q),\kappa^R(q)$ in Eq.~\eqref{eq:IsoGreensFunction} based on Eqs.~\eqref{eq:HarmonicFlexural} and \eqref{eq:HarmonicInPlane}. For isotropic systems that aren't under any stress, the correlation function of the in-plane phonons reduces to same form as Eq.~\eqref{eq:IsoGreensFunction}. However, as previously explained, in the presence of anisotropic stress the renormalized moduli may also become anisotropic.

We can now ask ourselves at what scale such a stress becomes important and thus induces anisotropy. Observing Eq.~\eqref{eq:EffectiveGreenFlexural}, it can be seen that under tensile stresses, one may note the introduction of a new length scale when $\sigma_{\alpha \alpha}q_{\alpha}^2 \sim B_{ijkl}^R(q)q_iq_jq_kq_l$ \cite{kovsmrlj2016response}. This length scale can be identified with the scale beyond which stress plays a dominant role along the axes in which it is present. For sufficiently small stresses such that $q_{\sigma} \ll q_{\text{th}}$ and assuming that the material is isotropic at $T=0$ and thus we can assume that $B_{ijkl}^R(q) \sim \kappa^R(q) \sim (q/q_{\text{th}})^{-\eta}$. Thus the wave vector takes the form \cite{kovsmrlj2016response}:
\begin{equation}
\label{eq:lengthscalerelationlow}
    q_{\sigma} = \bigg( \frac{\sigma}{\kappa} \bigg)^{1/(2-\eta)} q_{\text{th}}^{-\eta/(2-\eta)}
\end{equation} 
The value of stress for which these two length scales are equal can be solved for $\sigma_{q_{th}} = \kappa q_{\text{th}}^2= \kappa (\frac{4 \mu ( \lambda+\mu)k_BT}{\kappa^2( \lambda+2\mu)})^{\frac{1}{4-D}}$. For very large stress, $\sigma \gg \sigma_{q_{th}}$, where $q_{\sigma}  \gg q_{\text{th}}$, then anomalous behaviors will not enter into the comparison between the stress and bending portions of the flexural correlation function and thus the bare parameters can be used, $\sigma q^2 \sim \kappa q^4$, resulting in:
\begin{equation}
\label{eq:lengthscalerelationhigh}
   q_{\sigma} =  \bigg( \frac{\sigma}{\kappa}\bigg)^{1/2}
\end{equation}
With this pre-amble we may now begin to investigate the scaling theory of ``uni-axial'' stresses imposed upon thermalized elastic membranes.

\section{Scaling Behavior Of Elastic Moduli \label{sec:ScaleBehavior}}
In the following text, we aim to derive the scaling of the correlation functions in different regimes which will depend on the ordering of $q,q_{\sigma},q_{\text{th}}$ (note that order of these Fourier scales can change by tuning the parameters $\sigma,\kappa,T,Y$ as well as trivially changing the inverse length scale $q$). In order to develop a convention for naming these different regimes, we define un-ambiguously that a system under ``low'' stress to be such that $q>q_{\sigma}$ and ``high'' stress such that $q<q_{\sigma}$. We also define systems under ``low'' temperature conditions to be such that $q>q_{\text{th}}$ and high temperature such that $q<q_{\text{th}}$.

To obtain the scaling of correlation function we must commence with the calculation of the engineering dimensions when stress is ``low'' and when stress is ``high''. Engineering dimensions will tell us whether terms are relevant or irrelevant to the theory. Specifically, terms with negative engineering dimension are called irrelevant and can be ignored from the scaling theory of the free energy. On the other hand, terms with positive engineering dimensions are called relevant and cannot be ignored from the scaling of the theory. An-harmonic terms with positive engineering dimension can induce anomalous scaling of the elastic moduli of the theory that is different from the linear theory (such as those for the un-stressed isotropic elastic membranes in Eq.~\eqref{eq:IsoGreensFunction}). Thus it will be necessary to derive the Self-Consistent Screening Analysis (SCSA) equations, which allow us to describe the effect of an-harmonic terms in the free energy. With these two tools, we will then derive the scaling of the correlation functions in each regime.

\subsection{Engineering Dimensions}
\subsubsection{Low Stress \texorpdfstring{$q >  q_{\sigma}$}{TEXT}}
Before engineering dimensions are calculated, it is important to establish what the dominant term of the harmonic portion of the free energy is in order to proceed further into our scale-dependent analysis. In the presence of vanishingly small stresses, one can consider the non-anomalous correlation functions (also known as harmonic propagators) to scale as $\mathcal{G}_{ff} (q) \sim q^{-4}$, $\mathcal{G}_{uu} (q) \sim q^{-2}$ (as can be seen from Eqs.~\eqref{eq:HarmonicFlexural} and  \eqref{eq:HarmonicInPlane}), and see that the flexural modes fluctuate with a larger amplitude for small enough $q$ and thus produce the dominant modes of the harmonic portion of the free energy, otherwise known as the harmonic/Gaussian theory. In other words, the term $B_{ijkl} \partial_i \partial_j f \partial_k \partial_l f$ is the dominant term in the Gaussian theory.

Thus, in the presence of the scale-invariant dominant term, $B_{ijkl} \partial_i \partial_j f \partial_k \partial_l f$, we may calculate how fields $f$ should be re-scaled $f(\mathbf{q}) \equiv b^{- \Delta_f} f'(\mathbf{q}')$ as a result of the scale transformation $\mathbf{q} = b \mathbf{q}' \equiv b^{\Delta_q} \mathbf{q}' $ where $b>1$ is a rescaling parameter.

Engineering dimensions give the exponent with which parameters of a theory rescale (in this case $C_{ijkl},\sigma,B_{ijkl}$ though $B_{ijkl}$ will be scale-invariant since we set it to our dominant term), $O \equiv b^{-\Delta_O} O'$, under the scale transformation $q =b q'$. If an engineering dimension, $\Delta_O$, of a parameter is positive then it cannot be ignored as $q \rightarrow 0$ since it grows with scale. If on the other hand it is negative, then the parameter rescales to zero as $q \rightarrow 0$ and can thus be ignored (unless it is dangerously irrelevant) \cite{tongstatistical}.

\begin{table}
{
    \begin{tabular}{ |p{2cm}|p{2cm}|p{2.5cm}|  }
 \hline
 \multicolumn{2}{|c|}{$q >q_{\sigma} $, $\sigma_{\alpha \alpha}=\sigma,\sigma_{DD}=0$} \\
 \hline
 Term&  Eng. Dim. \\
 \hline
 $\Delta_{q }$   & $1$   \\
 $\Delta_{f }$   & $(4-D)/2$   \\
 $\Delta_{u}$ &   $3-D$     \\
 $\Delta_{C_{ijkl}}$ &   $4-D$  \\
 $\Delta_{B_{ijkl}}$ &   $0$\\
 $\Delta_{\sigma_{\alpha \alpha}}$ &   $2$ \\
 \hline
    \end{tabular}
}
\caption{\label{tab:tablow} In the presence of small ``uniaxial'' stress and high temperatures, engineering dimensions of the order parameters and the elastic moduli of the theory. }
\end{table}

Proceeding to the counting of engineering dimensions, at low stresses, the dominant term of our theory is $B_{ijkl} q_iq_jq_kq_lf(\mathbf{q})f(-\mathbf{q})$. This will automatically imply that the engineering dimensions $\Delta_{B_{ijkl}} = 0$. This implies then:
\begin{equation}
    \begin{split}
    & A \sum_{|\mathbf{q}|<\Lambda} B_{ijkl} \  q_iq_jq_kq_lf(\mathbf{q})f(-\mathbf{q}) \\  & = b^{-D} A' \sum_{|\mathbf{q}'|<\Lambda/b} B_{ijkl} \ b^{4} q'_iq'_jq'_kq'_l f(b\mathbf{q}')f(-b\mathbf{q}')
    \end{split}
\end{equation}
where $b^{-D}A' = A$ due to the fact that the area is a $D$-dimensional area in real space. In order for this term to be scale invariant then we must enforce that $b^{(4-D)/2} f(b\mathbf{q}') = f'(\mathbf{q}') $. Thus, we have obtained $\Delta_f = (4-D)/2$, and we can use this to obtain how the order parameter $u$ and all other coefficients of terms in the free energy re-scale. 

Due to the rotational symmetry of the free energy, the strain tensor will be preserved despite renormalization \cite{guitter1989thermodynamical, amit2005field}. Thus, the scale-invariance of the theory will enforce that the $u$ field rescales $u(\mathbf{q}) \equiv b^{-\Delta_u} u'(\mathbf{q}')$ in such a way that the individual terms of the strain tensor $q_i u_j $ and $q_i f q_j f$ also re-scale the same way. Thus, equating $q_i u_j \sim q_i f q_j f$ leads to
\begin{equation}
\begin{split}
    & q_i u_j(\mathbf{q}) = b q_i' u_j(b\mathbf{q}') \equiv b^{-\Delta_u+1}q_i'u_j'(\mathbf{q}') \sim \\
    & q_i q_j  f(\mathbf{q})f(-\mathbf{q}) = b^{2} q_i' q_j' f(b\mathbf{q}')f(-b\mathbf{q}') \\ & \equiv b^{-2 \Delta_f +2} q_i' q_j'f'(\mathbf{q}')f'(-\mathbf{q}')
\end{split}
\end{equation}

In other words, $\Delta_u-1 =   2\Delta_f-2 $ which gives $\Delta_u=3-D$. With this, we can finally calculate how the coefficients of an-harmonic terms, $C_{ijkl}$, of the free energy in Eq.~\eqref{eq:FreeEnergyAniso} should re-scale. For example, one can obtain:
\begin{equation}
    \begin{split}
        & A \sum_{|\mathbf{q}|<\Lambda} C_{ijkl}  \ q_iq_ku_j(\mathbf{q})u_l(-\mathbf{q})
        \\  & = b^{-D} A' \sum_{|\mathbf{q}'|<\Lambda/b} C_{ijkl} \ b^{2} q'_iq'_k b^{2D-6} u_j(b\mathbf{q}')u_l(-b\mathbf{q}')
    \end{split}
\end{equation}
which implies then that $C_{ijkl}' \equiv  C_{ijkl} b^{D-4}$ and gives us the engineering dimension $\Delta_C = 4-D$. Given these results then we know that when $D<4$, that an-harmonic terms with coefficients $C_{ijkl}$ will be relevant to physical behavior in the limit $q\rightarrow 0$. If stresses are not vanishingly small, a similar analysis can be done to show that $\Delta_{\sigma} = 2$ which indicates that it will be strongly relevant and that once $q<q_{\sigma}$, it can no longer be ignored. Thus the dominant term of the theory would have to be reconsidered in the ``high'' stress case which we will deal with in the very next section.
All engineering dimensions  for the ``low'' stress case are summarized in Table~\ref{tab:tablow}.

\subsubsection{High Stress \texorpdfstring{$q <  q_{\sigma}$}{TEXT} \label{subsec:HighStressEngDim}}
\begin{table}
{
\begin{tabular}{ |p{2cm}|p{2cm}|p{2.5cm}|  }
 \hline
 \multicolumn{2}{|c|}{$q <q_{\sigma},\sigma_{\alpha \alpha}>0,\sigma_{DD}=0$} \\
 \hline
 Term&   Eng. Dim.\\
 \hline
 $\Delta_{q_{\alpha}}$     &$2$\\
 $\Delta_{q_D}$     &$1$\\
 $\Delta_{f }$      &$(5-2D)/2$\\
 $\Delta_{u_{\alpha} }$   & $3-2D$   \\
  $\Delta_{u_D }$ & $4-2D$   \\
 $\Delta_{C_{\alpha \alpha \beta \beta}}$   & $1-2D$  \\
 $\Delta_{C_{\alpha \beta \alpha \beta}}$  & $1-2D$  \\
 $\Delta_{C_{\alpha D \alpha D}}$  & $3-2D$\\
 $\Delta_{C_{\alpha \alpha DD}}$  & $3-2D$\\
 $\Delta_{C_{DDDD}}$  & $5-2D$\\
 $\Delta_{B_{\alpha \alpha \beta \beta}}$  & $-4$\\
 $\Delta_{B_{\alpha \alpha DD}}$ & $-2$\\
 $\Delta_{B_{DDDD}}$ & $0$\\
 $\Delta_{\sigma_{\alpha \alpha }}$ & $0$\\
 \hline
\end{tabular} 
}

\caption{\label{tab:tabhigh} In the presence of a large ``uni-axial''  tension, engineering dimensions are shown for spatial dimensions, order parameters and the elastic moduli of the theory. }
\end{table}

 As previously mentioned, when stress is significant, the dominant portion of the Gaussian theory has to be reconsidered. A ``uni-axial'' stress term, $\sigma q_{\alpha}^2$, will dominate over the bending rigidities in the flexural correlation function along every axis except for the $D^{\text{th}}$ axis. Thus the new dominant term of the Gaussian theory becomes:
\begin{equation} 
   [B_{DDDD} \ q_D^4+\sigma_{\alpha \alpha} \ q_{\alpha}^2]f(\mathbf{q})f(-\mathbf{q})
\end{equation}
Rendering these free energy terms scale invariant requires axes co-linear with the ``uni-axial'' stress must be re-scaled such that $q_1 \sim...\sim q_{D-1} \sim q_D^2$ \cite{diehl2000critical, tauber2014critical, hornreich1975critical}. Thus if the wave vectors rescale as $(\mathbf{q}_{D-1},q_D) \equiv  (b^2 \mathbf{q}_{D-1}',b q_D')$ where $\mathbf{q}_{D-1} = (q_1,...,q_{D-1})$, then keeping the term $A \sum_{\mathbf{q}} [B_{DDDD}q_D^4+\sigma_{\alpha \alpha}q_{\alpha}^2] f^2$ scale invariant leads to 
\begin{equation}
    \begin{split}
        &A \sum_{|\mathbf{q}|<\Lambda} [B_{DDDD} \ q_D^4+\sigma_{\alpha \alpha} \ q_{\alpha}^2] f(\mathbf{q}_{D-1},q_D)f(-\mathbf{q}_{D-1},-q_D) \\ &= b^{1-2D} A' \sum_{|\mathbf{q}'|<\Lambda/b} [B_{DDDD} \ b^{4} {q_D'}^4+\sigma_{\alpha \alpha} b^{4} {q_{\alpha}'}^2]\ \times \\ & \quad \quad\quad\quad\quad\quad \quad f(b^2 \mathbf{q}_{D-1}', bq_D')f(-b^2 \mathbf{q}_{D-1}', -bq_D')
    \end{split}
\end{equation}
where $A = b^{1-2D}A'$ since $A =L_D\prod_{\alpha=1}^{D-1} L_{\alpha} = b^{-1} L_D' \prod_{\alpha=1}^{D-1} b^{-2} L_{\alpha}'$ (where $L_i$ are the system dimensions along axis $i$ and re-scale inverse to how $q_i$ re-scale). This equation thus implies the engineering dimension $\Delta_f = (5-2D)/2$. Observing the difference between the two terms $B_{DDDD}q_D^4$ and $B_{\alpha \alpha DD}q_{\alpha}^2q_D^2 f(\mathbf{q})^2$ and since $\Delta_{B_{DDDD}}=0$ and $q_{\alpha} \sim q_D^2$, we can conclude that $\Delta_{B_{\alpha \alpha DD}} = -2$. Likewise, $\Delta_{B_{\alpha \alpha \beta \beta}} = -4$. As was done in the previous section we can compare the terms within the strain tensor $q_i u_j \sim q_i f q_j f$ and we get that
\begin{equation}
\begin{split}
    & q_i u_j(\mathbf{q}) = b^{\Delta_{q_i}} q_i' u_j(b\mathbf{q}') \equiv b^{-\Delta_u+\Delta_{q_i}}q_i'u_j'(\mathbf{q}') \sim \\
    & q_iq_jf(\mathbf{q})f(-\mathbf{q}) = b^{\Delta_{q_i} + \Delta_{q_j}}q_i'q_j'f(b\mathbf{q}')f(-b\mathbf{q}') \\ & \equiv b^{-2 \Delta_f +\Delta_{q_i} + \Delta_{q_j}} q_i' q_j' f'(\mathbf{q}')f'(-\mathbf{q}')
\end{split}
\end{equation}
and thus $\Delta_{u_i} = 2 \Delta_{f}-\Delta_{q_i}$. Because of the difference in how $q_{\alpha}$ and $q_D$ re-scale, we obtain two different engineering dimensions for the in-plane displacement fields: $\Delta_{u_{\alpha}} = 3-2D$ and $\Delta_{u_D} = 4-2D$. Due to the anisotropic re-scaling of spatial dimensions, this causes $C_{ijkl}$ with differing indices to be re-scaled differently as well. As an example, consider $C_{\alpha \alpha \beta \beta}$:

\begin{equation}
    \begin{split}
        & A \sum_{|\mathbf{q}|<\Lambda} C_{\alpha \alpha \beta \beta} \ q_{\alpha}q_{\beta}u_{\alpha}(\mathbf{q}_{D-1},q_D)u_{\beta}(-\mathbf{q}_{D-1},-q_D)
        \\  & = b^{1-2D} A' \sum_{|\mathbf{q}'|<\Lambda/b} C_{\alpha \alpha \beta \beta} \ b^{4} q'_{\alpha}q'_{\beta}\  \times \\ &\quad \quad \quad\quad\quad b^{4D-6} u_{\alpha}(b^2 \mathbf{q}_{D-1}',b q_{D}')u_{\beta}(- b^{2}\mathbf{q}_{D-1}',-b q_{D}')
    \end{split}
\end{equation}
thus implying $\Delta_{C_{\alpha \alpha \beta \beta}} = 1-2D$. Similar analysis yields as well that $\Delta_{C_{\alpha \beta \alpha  \beta}} = 1-2D$. On the other hand, for $C_{\alpha D \alpha D}$:
\begin{equation}
    \begin{split}
        & A \sum_{|\mathbf{q}|:q<\Lambda} C_{\alpha D \alpha D}  \ q_{\alpha}q_{\alpha} \ u_D(\mathbf{q}_{D-1},q_D)u_D(-\mathbf{q}_{D-1},-q_D)
        \\  & = b^{1-2D} A' \sum_{|\mathbf{q}'|<\Lambda/b} C_{\alpha D \alpha D} \ b^{4} q'_{\alpha}q'_{\alpha}\  \times \\ &\quad \quad\quad\quad\quad b^{4D-8} u_D(b^{2}\mathbf{q}_{D-1}',b q_{D}')u_D(- b^{2}\mathbf{q}_{D-1}',-b q_{D}')
    \end{split}
\end{equation}
thus implying $\Delta_{C_{\alpha D \alpha D}} = 3-2D$. Similar conclusions can be drawn for $\Delta_{C_{\alpha \alpha DD}} = 3-2D$ whereas $\Delta_{C_{DDDD}} = 5-2D$ as summarized in Table~\ref{tab:tabhigh}.

Thus, we see that when $D > 5/2$, all $C_{ijkl}^R$ become irrelevant and thus an-harmonic terms do not contribute to the theory and can be ignored in the limit $q \rightarrow 0$. On the other hand, in the case of interest $D=2$, we can remove all irrelevant terms ($C_{1111}^R,C_{1122}^R,C_{1212}^R,B_{1111}^R,B_{1122}^R$) in the expression of the free energy as they only add technical complications and do not contribute to the qualitative scaling behavior. One can then integrate out the in-plane phonons, $u_i(\mathbf{r})$, with only the constant $C_{2222}^R$ present in the free energy since all other $C_{ijkl}^R$ are irrelevant. This can be done by means of the functional integral:
\begin{equation}
    e^{-\beta \mathcal{F}_{eff}} = \int \prod_i D[u_i(\mathbf{r})] e^{-\beta \mathcal{F}}
\end{equation}

Such an integration will cause all $f^4$ vertices to disappear and thus the effective free energy will take a simplified form:
\begin{equation} 
\label{eq:effectivetheoryB2222}
    \frac{\mathcal{F}_{eff}}{A} = \frac{1}{2} \sum_{|\mathbf{q}|<q_{\sigma}}[B_{2222}^R(q_{\sigma})q_2^4+\sigma_{11}q_1^2]f(\mathbf{q})f(-\mathbf{q})
\end{equation}
Observing this equation, one may note the absence of an-harmonic terms. This implies that $B_{2222}$ will no longer renormalize once the ``uni-axial'' stress is dominant. Thus the bending rigidity satisfies $B_{2222}^R(\mathbf{q}) = B_{2222}^R(q_{\sigma})$ (and more generally $B_{DDDD}$) and actually remains a constant when $q<q_{\sigma}$.

We will now calculate the SCSA equations corresponding to the theory so that we may later on calculate the potential anomalous scaling of elastic moduli due to the presence of relevant an-harmonic terms in the free energy.

\subsection{SCSA And \texorpdfstring{$\beta$}{TEXT} Equations \label{sec:SCRG}}
In order to derive anomalous exponents of the elastic moduli and more precise values for the cross-over scales $q_{\sigma}$ , $q_{\text{th}}$ where scaling of the correlation functions change, it is important to take a moment to calculate the SCSA equations that take into account the effect of an-harmonic terms into the calculation of effective elastic moduli.

Before proceeding to the derivation of the SCSA equations, we will take a brief aside to mention that one can integrate out all in-plane phonons to obtain an effective free energy for the flexural field. This will be necessary to obtain the SCSA equation for the flexural correlation function $\mathcal{G}_{ff}^R(\mathbf{q})$. For the purpose of obtaining useful expressions, we assume the general orthorhombic symmetry of the elastic tensors. Integrating out the in-plane phonons for general $D$ gives rise to a complicated coefficient of the $f^4$ vertex. However, the effective energy for flexural phonons under periodic boundary conditions in the presence of a general stress takes the following form in $D=2$ \cite{nelson2004statistical}:
\begin{widetext}
\begin{equation}
\label{eq:EffectiveFreeEnergyStress}
\begin{split}
     \frac{\mathcal{F}_{eff}}{A} =  & \frac{1}{2} \sum_{|\mathbf{q}|<\Lambda} [B_{ijkl} \ q_iq_jq_kq_l + \sigma_{ij}q_iq_j] f(\mathbf{q})f(-\mathbf{q}) \\ &+\frac{1}{8} \sum_{\substack{\mathbf{q}_1+\mathbf{q}_2 = \\ -\mathbf{q}_3-\mathbf{q}_4=\mathbf{q}\neq 0, |\mathbf{q}_i|_{i=1,...,4}<\Lambda}} q^4[q_{1i}P^T_{ij}(\mathbf{q})q_{2j}][q_{3i}P^T_{ij}(\mathbf{q})q_{4j}] \frac{N}{E(\mathbf{q})}  f(\mathbf{q}_1)f(\mathbf{q}_2)f(\mathbf{q}_3)f(\mathbf{q}_4)
\end{split}
\end{equation}
\end{widetext}
where $N$ and $E(\mathbf{q})$ are:
\begin{equation}
\label{eq:EquationEffectiveCoeff}
    \begin{split}
        N  = &C_{1212}[C_{1111}C_{2222}-C_{1122}^2] \\
        E(\mathbf{q}) =& \text{Det}[C_{ijkl}q_iq_k] = \\
        E(\mathbf{q}) =&C_{1111}C_{1212}q_1^4+C_{2222}C_{1212}q_2^4 \\
        &
        +(C_{1111} C_{2222}-2C_{1122}C_{1212}-C_{1122}^2)q_1^2q_2^2 \\
    \end{split}
\end{equation}

\begin{figure*}
\includegraphics{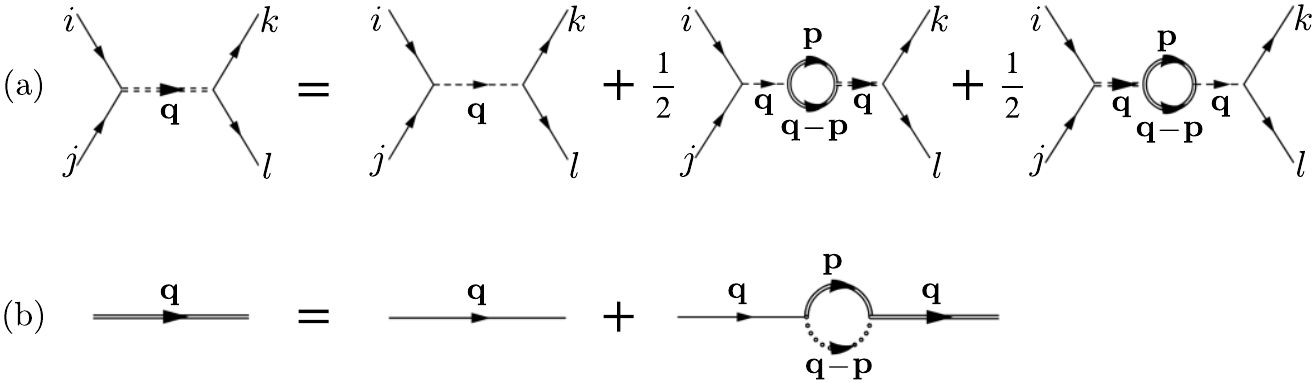}
\caption{(a) The SCSA equation is shown graphically using the $C_{ijkl}\partial_if\partial_jf\partial_kf\partial_lf$ vertex. This equation is used to obtain a scaling of the $C_{ijkl}$ via a self-consistent analysis. The symmetrization is due to the major symmetry of the Hamiltonian. The the dashed line indicates $C_{ijkl}$ and the doubled dashed line $C_{ijkl}^R$. The solid lines indicate $\mathcal{G}_{ff}$ whereas the doubled solid lines indicate $\mathcal{G}_{ff}^R$. (b) The SCSA equation corresponding to the flexural correlation function is shown using the effective $f^4$ vertex in Eq.~\eqref{eq:EffectiveFreeEnergyStress}. The renormalized structure of the vertex is marked by the doubled dotted line.}
\label{fig:Feynman1}
\end{figure*}

Returning now to the derivation of the SCSA equation, one can compute one-loop SCSA equations for the in-plane moduli. In principle one can do calculations to higher order loops to obtain more accurate results, however one can often gain a qualitative understanding from a one-loop analysis. From an SCSA shown in Fig.~\ref{fig:Feynman1} we can derive the following self-consistent equations for $C_{ijkl}^R$. 

\begin{widetext}
\begin{equation}
\label{eq:SCC}
    \begin{split}
     C_{ijkl}^R(\mathbf{q}) =  &C_{ijkl} \\ & - \frac{k_BT}{4(2 \pi )^D} \int_{|\mathbf{p}|<\Lambda} d^D \mathbf{p}    [C_{ijmn}^R(\mathbf{q})(q_m-p_m) p_n][C_{abkl}(q_a-p_a) p_b]\frac{A\mathcal{G}_{ff}^R(\mathbf{q}-\mathbf{p})}{k_BT}\frac{A\mathcal{G}_{ff}^R(\mathbf{p})}{k_BT} \\ & - \frac{k_BT}{4(2 \pi )^D} \int_{|\mathbf{p}|<\Lambda} d^D \mathbf{p}   [C_{ijmn}(q_m-p_m) p_n][C_{abkl}^R(\mathbf{q})(q_a-p_a) p_b]\frac{A\mathcal{G}_{ff}^R(\mathbf{q}-\mathbf{p})}{k_BT}\frac{A\mathcal{G}_{ff}^R(\mathbf{p})}{k_BT}
    \end{split}
\end{equation}
\end{widetext}

The symmetrization of the diagrammatic contributions seen in Fig.~\ref{fig:Feynman1} is due to the major symmetry of the tensor $C_{ijkl}=C_{klij}$ which enforces conservation of energy. This symmetry should remain even through renormalization. One can also obtain identical SCSA equations renormalizing the $C_{ijkl} \partial_i u_j \partial_kf \partial_l f$ vertex since the form of Hamiltonian will be preserved via renormalization \cite{guitter1989thermodynamical}. Similarly a self-consistent equation can be written down for the flexural correlation function:
\begin{widetext}
\begin{equation}
\label{eq:SCB}
    \mathcal{G}_{ff}^R(\mathbf{q}) = \mathcal{G}_{ff}(\mathbf{q})-\frac{A}{k_BT} \sum_{\mathbf{p}:p<\Lambda} q^4[p_{i}P^T_{ij}(\mathbf{q}-\mathbf{p})q_{j}]^2  \frac{N}{E(\mathbf{q}-\mathbf{p})} \mathcal{G}_{ff}^R(\mathbf{p}) \mathcal{G}_{ff}^R(\mathbf{q})\mathcal{G}_{ff}(\mathbf{q})
\end{equation}
\end{widetext}

One can also obtain the corresponding momentum shell Renormalization-Group equations, or $\beta$ equations, which will be of use to derive the values of $q_{\text{th}}$ and $q_{\sigma}$ more precisely \cite{altland2010condensed}. One can apply the operator, $-q \partial_q \equiv \partial_s$ to the SCSA equation~\eqref{eq:SCC}, and convert the Fourier sum to a momentum shell integral \cite{radzihovsky1991statistical,kovsmrlj2016response,wilson1974renormalization} in which we have integrate a momentum shell of Fourier space $\Lambda/b<p<\Lambda$ of the $\mathbf{p}$ integral: 
\begin{equation}
\label{eq:RGC2222}
\begin{split}
  & \partial_s C_{ijkl}^R(s)  = 2(2\Delta_f-1) C_{ijkl}^R(s) \\&\quad -  \frac{k_BT \Lambda^{D-4}}{2(2 \pi )^D}  \int d^{D-1}\hat{\mathbf{p}} 
     \frac{[C_{ijmn}^R(s)\hat{p}_m \hat{p}_n][C_{abkl}^R(s)\hat{p}_a \hat{p}_b] }{[B_{ijkl}^R(s)\hat{p}_i\hat{p}_j\hat{p}_k\hat{p}_l+\frac{\sigma_{ij}}{\Lambda^2}\hat{p}_i \hat{p}_j]^2} 
    \end{split}
\end{equation}
where $s$ is the rescaling parameter, with $b = e^s; 0<s \ll 1$. We have extracted the powers of $\Lambda$ and this leaves us with an angular integral over the unit vectors $\hat{p}$. Furthermore $\Delta_f$ is the engineering dimension of the order parameter $f$, though in renormalization, it is generally treated as a degree of freedom \cite{kovsmrlj2016response,kardar2007statistical,radzihovsky1991statistical}. Once the momenta in the window $\Lambda/b<p<\Lambda$ have been integrated over, the UV cutoff $\Lambda/b$ is re-scaled to $\Lambda$.

In the limit of vanishing stress one can additionally write down a general $D$-dimensional $\beta$ function for the isotropic bending rigidity $\kappa$ \cite{radzihovsky1991statistical}:
\begin{equation}
\begin{split}
\label{eq:kappaLth}
    \partial_s \kappa = ( 2\Delta_f - D - 4) \kappa  +  \frac{4[D^2-1]\mu(\lambda+\mu) S_D k_B T}{(D^2+2D)(\lambda+2\mu)(2 \pi)^D \kappa \Lambda^{4-D}}  
\end{split}
\end{equation}
where $S_D$ is the area of a $D$-dimensional unit sphere.

With the SCSA equations and $\beta$-functions calculated, we can now proceed to obtaining the scaling of correlation functions in each regime and obtain the scale limits of each regime. We order the regimes by investigating regimes with different dominant harmonic terms independently: in other words low stress, $q>q_{\sigma}$ and high stress $q<q_{\sigma}$. Within each of these regimes, we investigate each sub-regime depending on whether temperature is significant or not: in other words low temperature, $q>q_{\text{th}}$ and high temperature $q<q_{\text{th}}$.
\subsection{Scaling At Low Stress \texorpdfstring{$q >  q_{\sigma}$}{TEXT} \label{sec:LowStress}}

\subsubsection{Scaling At Low Temperature \texorpdfstring{$q > \max \{ q_{\sigma}, q_{\text{th}} \}$}{TEXT} }
The positive engineering dimension of $C_{ijkl}$ when $D<4$ signifies the importance of the parameter when $q \rightarrow 0$, however this does not mean that an-harmonic effects need to be considered at smaller finite length scales.

Indeed, when $q > \max \{q_{\sigma} , q_{\text{th}}\}$, the scaling of the correlation functions is trivial since both temperature and stress do not contribute significantly, thus the harmonic low stress correlation functions are sufficient to understand the theory and no anomalous effects should be observed in the theory. Reminding the reader once again that we are assuming the Roman letter indices such as $i,j,k,l$ range over $\{ 1,...,D \}$ and Greek letter indices range over $\{ 1,...,D-1\} $ we have:
\begin{equation}
\begin{split}
    \mathcal{G}_{ff}(\mathbf{q}) = & \frac{k_BT}{A[B_{ijkl}q_iq_jq_kq_l +\sigma_{\alpha \alpha }q_{\alpha}^2]}\\  \approx & \frac{k_BT}{A[B_{ijkl}q_iq_jq_kq_l ]}-\frac{k_BT\sigma_{\alpha \alpha}q_{\alpha}^2}{A[B_{ijkl}q_iq_jq_kq_l ]^2}  
\end{split}
\end{equation}
where a Taylor expansion was performed if one is curious of the linear response of the theory and all $B_{ijkl}$ are the bare parameters of the theory since no thermal anomalous effects are relevant. Assuming the bare material is isotropic, we then have:
\begin{equation}
\begin{split}
    \mathcal{G}_{ff}(\mathbf{q}) \approx & \frac{k_BT}{A \kappa q^4}
\end{split}
\end{equation} 
where the stress term is insignificant to the scaling of the flexural correlation function.
And for the in-plane phonons we have:
\begin{equation}
\mathcal{G}_{u_i u_j} = \frac{k_BT}{A} [C_{ikjl}q_kq_l]^{-1}
\end{equation}
where the $C_{ijkl}$ are also the bare parameters of the theory.

Assuming a vanishing stress such that $q_{\sigma}  \ll q_{\text{th}}<q$, one can ask at what $q_{\text{th}}$ these harmonic approximations are no longer appropriate to use and thus an-harmonic terms play an important role. To do this, it is necessary to resort to the $\beta$ equations presented in Sec.~\ref{sec:SCRG}. Specifically we can examine Eq.~\eqref{eq:kappaLth} and look at what re-scaled UV cutoff $\Lambda_\text{th}$ the an-harmonic contribution is of the order of $\kappa$. Assuming isotropy and ignoring infinitesimal stresses, we have \cite{radzihovsky1991statistical}:
\begin{equation}
    \kappa \approx  \frac{4(D^2-1)k_BT \mu(\lambda+\mu) S_D \Lambda_{\text{th}}^{D-4}}{(\lambda+2\mu) (D^2+2D) \kappa (2 \pi )^D} 
\end{equation}
where $S_D$ is the area of a unit $D$-dimensional sphere. Therefore we obtain:
\begin{equation}
\label{eq:ThermalLength}
    q_{\text{th}} \equiv \Lambda_{\text{th}} = \bigg(4 \frac{(D^2-1)k_BT \mu(\lambda+\mu) S_D}{(\lambda+2\mu)(D^2+2D) (2\pi)^D  \kappa^2} \bigg)^{\frac{1}{4-D}}
\end{equation}
Likewise, assuming very small temperatures such that $q_{\text{th}} \ll q_{\sigma}<q$, one can similarly ask at what $q_{\sigma}$ the Gaussian theory breaks down due to stress. The answer to this is already in our derivation of Eq.~\eqref{eq:lengthscalerelationhigh} for $q_{\sigma}$. When $q_{\text{th}}<q<q_{\sigma}$, the dominant terms in the harmonic theory will have to be re-examined, which will be done in the following Sec.~\ref{sec:HighStress}. 
\subsubsection{Scaling At High Temperature \texorpdfstring{$q_{\text{th}}>q >  q_{\sigma} $}{TEXT}}

Having obtained the engineering dimensions we can make certain inferences of the behavior of these elastic sheets beyond $q_{\text{th}}$. Indeed when $q>q_{\sigma}$, the engineering dimensions seen in Table~\ref{tab:tablow} indicate that all the in-plane elastic moduli have positive engineering dimensions and must thus be considered in the theory. Indeed these are the same engineering dimensions as were found in \cite{guitter1989thermodynamical}. One can then perform an SCSA or renormalization group analysis \cite{le1992self,guitter1989thermodynamical,nelson1987fluctuations} to obtain the anomalous exponents assuming negligible stress. Thus, the non-trivial scaling analysis of stress-free thermally fluctuating membranes will hold \cite{kovsmrlj2016response}. Thus, assuming a large separation of the three Fourier scales and $q_{\sigma} \ll q \ll q_{\text{th}}$ and $D<4$, we expect to observe the anomalous thermalized exponents $\eta,\eta_u$ since the bending modes are dominant and the stress term is not significant. Thus we have:
\begin{equation}
\begin{split}
    \mathcal{G}_{ff}(\mathbf{q}) = & \frac{k_BT}{A[B_{ijkl}^R(\mathbf{q})q_iq_jq_kq_l +\sigma_{\alpha \alpha}q_{\alpha}^2 ]} \\ \approx &\frac{k_BT}{A[B_{ijkl}^R(\mathbf{q})q_iq_jq_kq_l]} 
\end{split}
\end{equation}
where $B_{ijkl}^R(\mathbf{q})$ can be taken to be the isotropic $\kappa_R(\mathbf{q})$ assuming the material is isotropic at $T=0$ and thus $\kappa_R(\mathbf{q}) \sim (q/q_{\text{th}})^{-\eta}$. Stress is still not significant in its contributions to the flexural correlation functions and for this reason it can be Taylor expanded. Analogously, 
\begin{equation}
\mathcal{G}_{u_i u_j} = \frac{k_BT}{A} [C_{ikjl}^R(\mathbf{q})q_kq_l]^{-1}
\end{equation}
and $C_{ikjl}^R(\mathbf{q}) \sim (q/q_{\text{th}})^{\eta_u}$. The scaling of the elastic moduli can be observed in Table~\ref{tab:tabscaling}.

Assuming, $q_{\sigma}<q<q_{\text{th}}$ we can ask ourselves once more, up to what $q_{\sigma}$ will this non-trivial scaling hold. To do this we merely repeat the steps indicated to derive Eq.~\eqref{eq:lengthscalerelationlow}. Thus we are aware of what is the range of this scaling regime in the presence of small stresses and must always be sure to use the anomalous exponents $\eta,\eta_u$ only when $q_{\sigma}<q<q_{th}$.

\subsection{Scaling At High Stress \texorpdfstring{$q <  q_{\sigma}$}{TEXT} \label{sec:HighStress}}

\subsubsection{Scaling At Low Temperature \texorpdfstring{$ q_{\text{th}}<q <q_{\sigma}   $}{TEXT}}
The stress length scale of this regime, establishing one of the bounds, is once again given by Eq.~\eqref{eq:lengthscalerelationhigh}. Since we are considering the case of low temperature, the renormalizing effect of an-harmonic terms can be ignored. Since no anomalous behaviors are expected, the flexural correlation function when stress is ``uni-axial'' is easily written down as:
\begin{equation}
\begin{split}
    \mathcal{G}_{ff}(\mathbf{q}) \approx & \frac{k_BT}{A[B_{DDDD}q_D^4 +\sigma_{\alpha \alpha}q_{\alpha}^2 ]} 
\end{split}
\end{equation}
where $B_{DDDD}$ is the $T=0$ value (which is once again $\kappa$ for an isotropic material).
Similarly, the in-plane phonon correlation functions should not show any anomalous behavior and we should observe:
\begin{equation}
\mathcal{G}_{u_i u_j}(\mathbf{q}) = \frac{k_BT}{A} [C_{ikjl}q_kq_l]^{-1}
\end{equation}
with the $T=0$ parameters of the theory being used. These scaling laws can be observed in Table~\ref{tab:tabscalinglsiglessthanlth}.

In the high stress case, we can once again ask where is the breakdown of the harmonic theory, in other words the new value of $q_{\text{th}}$. It need not be the same as the formula for the low stress case in Eq.~\eqref{eq:ThermalLength} due to stress now being significant. However, for $D>2$, there is no breakdown of the harmonic theory since all an-harmonic terms in the free energy are irrelevant. A length scale where the harmonic theory breaks down can be evaluated for $D=2$ for which $C_{DDDD}=C_{2222}$ is a relevant parameter. Thus we will calculate the value of $q_{\text{th}}$ explicitly in the case of $D=2$.

From the engineering dimensions, we understood that even beyond $q_{\text{th}}$, the flexural phonons should not show any anomalous behavior since $B_{DDDD}^R=B_{2222}^R$ will remain a constant. Indeed the only relevant parameter that can show anomalous behavior is $C_{DDDD}^R=C_{2222}^R$. Thus we repeat our derivation of $q_{\text{th}}$ in a similar manner as the derivation of Eq.~\eqref{eq:ThermalLength}, except we use the $\beta$ equation of $C_{2222}^R$ given in Eq.~\eqref{eq:RGC2222}:
\begin{widetext}
\begin{equation}
    \partial_sC_{2222}^R =2C_{2222}^R- \frac{k_BT}{2(2 \pi \Lambda_{th})^2}  \int_{0}^{2 \pi} d\theta 
     \frac{[C_{2222}^R\sin^2{\theta}+C_{1122}^R\cos^2{\theta}]^2 }{[B_{ijkl}^R(s)\hat{p}_i\hat{p}_j\hat{p}_k\hat{p}_l+\frac{\sigma_{11}}{\Lambda_{th}^2}\cos^2{\theta}]^2}
\end{equation}
\end{widetext}
where $\Delta_f=1$. Anomalous effects cannot be ignored once $\Lambda$ takes on a value such that the two terms on the right hand side are of the same order. Here we have once again named this inverse length scale as $\Lambda_{th}$. In other words we are interested to know at what scale the anomalous contribution becomes significant with respect to the linear term in the $\beta$ equation:
\begin{equation}
    2C_{2222}^R \approx \frac{k_BT}{2(2 \pi \Lambda_{th})^2}  \int_{0}^{2 \pi} d\theta 
     \frac{[C_{2222}^R\sin^2{\theta}+C_{1122}^R\cos^2{\theta}]^2 }{[B_{ijkl}^R(s)\hat{p}_i\hat{p}_j\hat{p}_k\hat{p}_l+\frac{\sigma_{11}}{\Lambda_{th}^2}\cos^2{\theta}]^2}
\end{equation}
We can approximately rewrite the denominator with the bare parameters as:
\begin{equation}
    B_{ijkl}^R\hat{p}_i\hat{p}_j\hat{p}_k\hat{p}_l+\frac{\sigma_{11}}{\Lambda_\text{th}^2}\cos^2{\theta} \longrightarrow \kappa +\frac{\sigma_{11}}{\Lambda_\text{th}^2}\cos^2{\theta}
\end{equation}
where we set $\kappa = B_{1111} = B_{2222} = B_{1122}$ as the bare isotropic bending rigidity. This is justified since an-harmonic contributions are insignificant and thus the bare bending rigidities can be used. Replacing $C_{1122}^R$ and $C_{2222}^R$ by their bare isotropic values $\lambda, \lambda+2\mu$ respectively, one can then perform the integral identity holds when $q$ is less than:
\begin{equation}
    \Lambda_{\text{th}} \equiv q_{\text{th}} = \frac{k_BT (\lambda+2\mu)}{16 \pi \sqrt{\kappa^3 \sigma}}
\end{equation}
Thus in the large stress limit we have a different form for $q_{\text{th}}$ which still matches with the low stress limit formula, Eq.~\ref{eq:ThermalLength}, for $q_{\text{th}}$ when $q_{\text{th}} \approx q_{\sigma}$ and $D=2$.

\subsubsection{Scaling At High Temperature \texorpdfstring{$q<\min \{ q_{\sigma} , q_{\text{th}}\} $}{TEXT} for \texorpdfstring{$D=2$}{TEXT}}


As was shown in Sec.~\ref{subsec:HighStressEngDim}, we know that below $q_{\sigma}$, $B_{DDDD}^R$ should not be anomalous and should be a constant at some finite value, we can use the SCSA equations to obtain the scaling behavior of the moduli in the long wavelength limit when $q<\min \{q_{\sigma},q_{\text{th}} \}$. Such an analysis is done in \cite{radzihovsky1995new} for tubules as well for general $D$.

However, we again restrict our focus to the scaling analysis of the elastic moduli in strictly $D=2$. This is not only the physically interesting case but also the least trivial due to the relevance of $C_{2222}$ for $D=2$ (whereas for larger dimensions all $C_{ijkl}$, including $C_{DDDD}$, become irrelevant as $q \rightarrow 0$). Furthermore, the analysis of the SCSA equations will be technically clearer in $D=2$.

We can now proceed to the SCSA using the same idea of removing irrelevant an-harmonic coefficients. We shall further assume that we have integrated all high-frequency modes with $q>\min\{q_{\sigma},q_{\text{th}}\}$ and thus all un-integrated wave-vectors in the following analysis satisfy $q<\min\{q_{\sigma},q_{\text{th}}\}$.

 Before beginning the analysis we note once again from Table~\ref{tab:tabhigh} that when $q<\min\{q_{\sigma},q_{\text{th}}\}$ all bending rigidities except for $B_{2222}$ are irrelevant. In addition $B_{2222}^R$ becomes a constant so if we define $q_{\text{min}} \equiv \min\{q_{\sigma},q_{\text{th}}\}$ then we can approximate the correlation function for $q<q_{\text{min}}$ as:
\begin{equation}
\label{eq:GreensFlexuralApprox}
\begin{split}
     \mathcal{G}_{ff}^R(\mathbf{q})  & =  \frac{k_BT}{A[B_{ijkl}^R(\mathbf{q})q_iq_jq_kq_l +\sigma_{11}q_1^2]} \\ & \approx \frac{k_BT}{A[B_{2222}^R(\mathbf{q})q_2^4 +\sigma_{11}q_1^2]}\\ & \approx \frac{k_BT}{A[B_{2222}^R(q_{\text{min}})q_2^4 +\sigma_{11}q_1^2]}
\end{split}
\end{equation}

As a final preliminary step we use a one loop SCSA analysis to obtain an-harmonic corrections to the elastic moduli resulting in Eq.~\eqref{eq:SCC}. We will divide this equation by $C_{ijkl}C_{ijkl}^R(\mathbf{q})$ and use the Eq.~\ref{eq:GreensFlexuralApprox} as the correlation functions leading to:
\begin{widetext}
\begin{equation}
\label{eq:SCCInv}
    \begin{split}
     \frac{1}{C_{ijkl}} =& \frac{1}{C_{ijkl}^R(\mathbf{q})}  \\ & - \frac{k_BT}{4(2 \pi )^2} \int_{|\mathbf{p}|<q_{\text{min}}} d p_1 dp_2   \frac{[C_{ijmn}^R(\mathbf{q})(q_m-p_m) p_n][C_{abkl}(q_a-p_a) p_b]}{C_{ijkl}C_{ijkl}^R(\mathbf{q})[B_{2222}^R(q_{\text{min}})(p_2-q_2)^4+\sigma_{11}(p_1-q_1)^2][B_{2222}^R(q_{\text{min}})p_2^4+\sigma_{11}p_1^2]}\\ & - \frac{k_BT}{4(2 \pi )^2} \int_{|\mathbf{p}|<q_{\text{min}}} d p_1 dp_2  \frac{[C_{ijmn}(q_m-p_m) p_n][C_{abkl}^R(\mathbf{q})(q_a-p_a) p_b]}{C_{ijkl}C_{ijkl}^R(\mathbf{q})[B_{2222}^R(q_{\text{min}})(p_2-q_2)^4+\sigma_{11}(p_1-q_1)^2][B_{2222}^R(q_{\text{min}})p_2^4+\sigma_{11}p_1^2]},
    \end{split}
\end{equation}
\end{widetext}
where $\mathbf{p} = (p_1,p_2)$. In the analysis we will drop the bounds of the integral and take it as a given that $|\mathbf{p}|<q_{\text{min}}$. We can now examine the scaling behavior of the elastic moduli as $q \rightarrow 0$. However, before beginning, it should be stated that though these results are derived from a one-loop SCSA, that the scalings for the elastic moduli in the rest of this section will be correct to all loops \cite{radzihovsky1995new}.

\paragraph{Scaling Behavior of $C_{2222}$.}

For $C_{2222}$, the corresponding self-consistent perturbative equation is given by Eq.~\eqref{eq:SCCInv} as:
\begin{widetext}
\begin{equation}
\label{eq:SCC22221}
\begin{split}
  \frac{1}{C_{2222}}  = \frac{1}{C_{2222}^R(\mathbf{q})} -  \frac{k_BT}{2(2 \pi)^2}  \int dp_1  dp_2 
     \frac{[(p_2-q_2)p_2+\frac{C_{1122}^R(\mathbf{q})}{C_{2222}^R(\mathbf{q})}(p_1-q_1)p_1][(p_2-q_2)p_2+\frac{C_{1122}}{C_{2222}}(p_1-q_1)p_1]   }{[B_{2222}^R (q_{\text{min}}) p_2^4+\sigma_{11}p_1^2][B_{2222}^R(q_{\text{min}})(p_2-q_2)^4+\sigma_{11}(p_1-q_1)^2]}
    \end{split}
\end{equation}
\end{widetext}
where $C_{2222}$ and $C_{1122}$ are the bare un-renormalized moduli. We examine Eq.~\eqref{eq:SCC22221} in the long wavelength limit when $q \rightarrow 0$, where $q<\min\{ q_{\sigma}, q_{\text{th}} \}$. From Table~\ref{tab:tabhigh} we know that $q_1 \sim q_2^2$ and $C_{1111},C_{1122},C_{1212},B_{1111},B_{1122}$ are all irrelevant.

We can then extract powers of $q_2$:

\begin{widetext}
\begin{equation}
\label{eq:SCC2222-11}
\begin{split}
   \frac{1}{C_{2222}}  = \frac{1}{C_{2222}^R(\mathbf{q})} -  \frac{k_BT}{2(2 \pi)^2}  \int d\tilde{p}_1  d\tilde{p}_2  \ q_2^3
     \frac{[\frac{1}{q_2^2}(\tilde{p}_2-1)\tilde{p}_2+\frac{C_{1122}^R(\mathbf{q})}{C_{2222}^R(\mathbf{q})}(\tilde{p}_1-\tilde{q}_1)\tilde{p}_1][\frac{1}{q_2^2}(\tilde{p}_2-1)\tilde{p}_2+\frac{C_{1122}}{C_{2222}}(\tilde{p}_1-\tilde{q}_1)\tilde{p}_1]   }{[B_{2222}^R(q_{\text{min}}) \tilde{p}_2^4+\sigma_{11}\tilde{p}_1^2][B_{2222}^R(q_{\text{min}})(\tilde{p}_2-1)^4+\sigma_{11}(\tilde{p}_1-\tilde{q}_1)^2]}
    \end{split}
\end{equation}

where $\tilde{p}_2=p_2/q_2$, $\tilde{p}_1=p_1/q_2^2$ and $\tilde{q}_1=q_1/q_2^2$. We must collect the most divergent terms as $q_2 \rightarrow 0$ since the left hand side of the equation is a constant. Since $C_{1122}$ is irrelevant we may also remove this term from the numerator and thus obtain the following equation:

\begin{equation}
\label{eq:SCC2222-111}
   \frac{1}{C_{2222}}  = \frac{1}{C_{2222}^R(\mathbf{q})} -  \frac{k_BT}{2(2 \pi)^2}  \frac{1}{q_2}\int  d\tilde{p}_1  d\tilde{p}_2  \frac{(\tilde{p}_2-1)^2 \tilde{p}_2^2 }{[B_{2222}^R(q_{\text{min}}) \tilde{p}_2^4+\sigma_{11}\tilde{p}_1^2][B_{2222}^R(q_{\text{min}})(\tilde{p}_2-1)^4+\sigma_{11}(\tilde{p}_1-\tilde{q}_1)^2]} 
\end{equation}
\end{widetext}

In the above equation, one can easily see that the integral is a homogeneous function of $\tilde{q}_1$, thus the self-consistent equation is solved by the ansatz:
\begin{equation}
    C_{2222}^R(\mathbf{q}) \approx C_{2222}^R(q_{\text{min}})\frac{q_2}{q_{\text{min}}} \Omega_C^2(1/\sqrt{\tilde{q}_1}) 
\end{equation}
where $\Omega_C^2(1/\sqrt{\sqrt{q}_1})$ is a universal scaling function that is a constant when $\tilde{q}_1 \rightarrow 0$. The pre-factor $C_{2222}^R(q_{\text{min}}) q_2 / q_{\text{min}}$ is meant to ensure that the correlation functions for $q<q_{\text{min}}$ and $q>q_{\text{min}}$ transition smoothly at $q_{\text{min}}$. The full form of $\Omega_C^2$ can be determined from the fact that $C_{2222}^R(\mathbf{q})$ should be independent of $q_1$ when $q_1 \rightarrow 0$ and it should be independent of $q_2$ when $q_2 \rightarrow 0$. Thus:
\begin{equation}
    \Omega_C^2(s) \sim 
     \begin{cases} 
      \text{constant}  & s \rightarrow \infty \\
      s^{-1} & s \rightarrow 0
   \end{cases}
\end{equation}
Assembling the pieces together we obtain:
\begin{equation}
\label{eq:C2222}
     C_{2222}^R(\mathbf{q}) \sim 
     \begin{cases} 
      q_2  & q_2 \gg \sqrt{q_1} \\
      \sqrt{q_1} & q_2 \ll \sqrt{q_1}
   \end{cases}
\end{equation}
Therefore, despite the fact that the effective theory in Eq.~\ref{eq:effectivetheoryB2222} does not possess anomalous scaling, the full theory with in-plane phonons does have anomalous exponents due to thermal fluctuations. The significance of this intuitively is that sinusoidal waves can form transverse to the axis of stress and are not flattened out by it. 

In addition, this result will be correct to all loops since one can check that at higher orders, the leading contribution to the SCSA equation of $C_{2222}^R$ at every order is $q_2 \Omega_C^2(q_2/\sqrt{q_1})$ \cite{radzihovsky1995new}.

\paragraph{Scaling Behavior of $C_{1122}$ and $C_{1111}$.}
Although the remaining moduli are irrelevant, we can repeat this same analysis to obtain how they scale. We can check for example how $C_{1122}^R$ and $C_{1111}^R$ should scale.

 We can use Eq.~\eqref{eq:SCCInv} and keep in mind that if we look at Table~\ref{tab:tabhigh}, we see that $C_{1111}^R$ is more irrelevant with respect to $C_{1122}^R$ and thus we can omit the $C_{1111}$ contributions to the SCSA equation of $C_{1122}^R$. This will result in: 
 \begin{widetext}
\begin{equation}
\label{eq:SCC1122-1}
    \begin{split}
   \frac{1}{C_{1122}}  =  \frac{1}{C_{1122}^R(\mathbf{q})} -  \frac{k_BT}{4 (2 \pi)^2} \int dp_1  dp_2 
    & \bigg[ \frac{ (p_2-q_2)p_2 [(p_1-q_1)p_1+\frac{C_{2222}^R(\mathbf{q})}{C_{1122}^R(\mathbf{q})} (p_2-q_2)p_2]  }{[B_{2222}^R(q_{\text{min}})p_2^4+\sigma_{11}p_1^2][B_{2222}^R(q_{\text{min}})(p_2-q_2)^4+\sigma_{11}(p_1-q_1)^2]} \\ &
     + \frac{ (p_2-q_2)^2 [p_1^2+\frac{C_{2222}}{C_{1122}} p_2^2]  }{[B_{2222}^R(q_{\text{min}}) p_2^4+\sigma_{11}p_1^2][B_{2222}^R(p_2-q_2)^4+\sigma_{11}(p_1-q_1)^2]} \bigg]
    \end{split}
\end{equation}
and
\begin{equation}
\label{eq:SCC1111-1}
\begin{split}
   & \frac{1}{C_{1111}}  = \frac{1}{C_{1111}^R(\mathbf{q})} -  \frac{k_BT}{2(2 \pi)^2}  \int dp_1  dp_2 
     \frac{[\frac{C_{1122}^R(\mathbf{q})}{C_{1111}^R(\mathbf{q})}(p_2-q_2)p_2+(p_1-q_1)p_1][\frac{C_{1122}}{C_{1111}}(p_2-q_2)p_2+(p_1-q_1)p_1]  }{[B_{2222}^R(q_{\text{min}})(p_2-q_2)^4+\sigma_{11}(p_1-q_1)^2][B_{2222}^R(q_{\text{min}})p_2^4+\sigma_{11}p_1^2]}
    \end{split}
\end{equation}

Similarly as before we can extract powers of $q_2$, keeping in mind also that $C_{2222}^R(\mathbf{q}) \sim q_2$. Thus giving:

\begin{equation}
\label{eq:SCC1122-11}
    \begin{split}
        \frac{1}{C_{1122}}  =  \frac{1}{C_{1122}^R(\mathbf{q})} -  \frac{k_BT}{4 (2 \pi)^2} \int \frac{d\tilde{p}_1  d\tilde{p}_2}{q_2^5} 
     & \bigg[ \frac{ (\tilde{p}_2-1)\tilde{p}_2q_2^2 [\tilde{p}_1(\tilde{p}_1-\tilde{q}_1)q_2^4+\frac{C_{2222}^R(q_{\min})\Omega_C^2(\frac{1}{\sqrt{\tilde{q}_1}})}{q_{\min}C_{1122}^R(\mathbf{q})} (\tilde{p}_2-1) \tilde{p}_2 q_2^3]  }{[B_{2222}^R(q_{\text{min}})\tilde{p}_2^4+\sigma_{11}\tilde{p}_1^2][B_{2222}^R(q_{\text{min}})(\tilde{p}_2-1)^4+\sigma_{11}(\tilde{p}_1-\tilde{q}_1)^2]} \\ &
     +\frac{ (\tilde{p}_2-1)\tilde{p}_2q_2^2 [(\tilde{p}_1-\tilde{q}_1)\tilde{p}_1q_2^4+\frac{C_{2222}}{C_{1122}} (\tilde{p}_2-1)\tilde{p}_2q_2^2]  }{[B_{2222}^R(q_{\text{min}}) \tilde{p}_2^4+\sigma_{11}\tilde{p}_1^2][B_{2222}^R(q_{\text{min}})(\tilde{p}_2-1)^4+\sigma_{11}(\tilde{p}_1-\tilde{q}_1)^2]}\bigg]
    \end{split}
\end{equation}

and

\begin{equation}
\label{eq:SCC1111-11}
\begin{split}
    \frac{1}{C_{1111}}  = \frac{1}{C_{1111}^R(\mathbf{q})} -  \frac{k_BT}{2(2 \pi)^2}  \int \frac{d\tilde{p}_1  d\tilde{p}_2}{q_2^5} 
     \frac{[\frac{C_{1122}^R(\mathbf{q})}{C_{1111}^R(\mathbf{q})}(\tilde{p}_2-1)\tilde{p}_2q_2^2+(\tilde{p}_1-\tilde{q}_1)\tilde{p}_1q_2^4][\frac{C_{1122}}{C_{1111}}(\tilde{p}_2-1)\tilde{p}_2q_2^2+(\tilde{p}_1-\tilde{q}_1)\tilde{p}_1q_2^4]  }{[B_{2222}^R(q_{\text{min}})(\tilde{p}_2-1)^4+\sigma_{11}(\tilde{p}_1-\tilde{q}_1)^2][B_{2222}^R(q_{\text{min}})\tilde{p}_2^4+\sigma_{11}\tilde{p}_1^2]}
    \end{split}
\end{equation}
\end{widetext}
By only paying attention to powers of $q_2$ and replacing finite integrals with the symbols $I^{(i)}$ we may write that:
\begin{equation}
    \begin{split}
        \frac{1}{C_{1122}} = & \frac{1}{C_{1122}^R(\mathbf{q})} +q_2I_{1122}^{(1)} +\frac{1}{C_{1122}^R(\mathbf{q})}I_{1122}^{(2)}  + \frac{1}{q_2}I_{1122}^{(3)}
    \end{split}
\end{equation}
and
\begin{equation}
    \begin{split}
        \frac{1}{C_{1111}} = & \frac{1}{C_{1111}^R(\mathbf{q})} +\frac{C_{1122}^R(\mathbf{q})}{C_{1111}^R(\mathbf{q})}\frac{1}{q_2}I_{1111}^{(1)} +\frac{C_{1122}^R(\mathbf{q})}{C_{1111}^R(\mathbf{q})}q_2I_{1111}^{(2)} \\ & 
        +q_2I_{1111}^{(3)} +q_2^3 I_{1111}^{(4)}
    \end{split}
\end{equation}
where the integrals $I_{1122}^{(i)}$ and $I_{1111}^{(i)}$ can be found in the appendix in Sec.~\ref{sec:Integrals}.
Collecting the most divergent powers (divergent as $q_2 \rightarrow 0$) in each equation gives:
\begin{equation}
\label{eq:C1122Scale}
    \begin{split}
        \frac{1}{C_{1122}} \approx & \frac{1}{C_{1122}^R(\mathbf{q})}(1+ I_{1122}^{(2)}) 
        + \frac{1}{q_2}I_{1122}^{(4)}
    \end{split}
\end{equation}
\begin{equation}
\label{eq:C1111Scale}
    \begin{split}
        \frac{1}{C_{1111}} \approx & \frac{1}{C_{1111}^R(\mathbf{q})} +\frac{C_{1122}^R(\mathbf{q})}{C_{1111}^R(\mathbf{q})}\frac{1}{q_2}I_{1111}^{(1)}
    \end{split}
\end{equation}
Eq.~\eqref{eq:C1122Scale} directly shows that it can be solved by the ansatz:
\begin{equation}
C_{1122}^R(\mathbf{q}) \approx C_{1122}^R(q_{\text{min}})\frac{q_2}{q_{\min}} \Omega_C^1(q_2/\sqrt{q_1}) 
\end{equation}
where we have once again a pre-factor that ensures the the correlation functions for $q<q_{\min}$ and $q>q_{\min}$ match. And the homogeneous function has the following scaling:
\begin{equation}
    \Omega_C^1(s) \sim 
     \begin{cases} 
      \text{constant}  & s \rightarrow \infty \\
      s^{-1} & s \rightarrow 0
   \end{cases}
\end{equation}
 and hence, $C_{1122}(\mathbf{q}) \sim q_2$ and thus that:
\begin{equation}
\label{eq:C1122}
     C_{1122}^R(\mathbf{q}) \sim 
     \begin{cases} 
      q_2  & q_2 \gg \sqrt{q_1} \\
      \sqrt{q_1} & q_2 \ll \sqrt{q_1}
   \end{cases}
\end{equation}
When we insert this ansatz into Eq.~\eqref{eq:C1111Scale} we obtain also that $C_{1111}^R$ becomes a constant and thus must be approximately $C_{1111}^R(q_{\text{min}}) $. These can be found in Tables~\ref{tab:tabscaling} and \ref{tab:tabscalinglsiglessthanlth}.

\paragraph{Scaling Behavior of $C_{1212}$.}
Lastly, we can check how the shear modulus should scale via its corresponding self-consistent equation:
\begin{equation}
    \begin{split}
    \frac{1}{C_{1212}}  = &\frac{1}{C_{1212}^R(\mathbf{q})} \\
    &-  \frac{2 k_BT}{(2 \pi)^2} \int dp_1  dp_2
     \frac{  p_1 p_2 }{[B_{2222}^R(q_{\text{min}})p_2^4+\sigma_{11}p_1^2]} \\&  \quad   \quad \times \frac{(p_1-q_1)(p_2-q_2)}{[B_{2222}^R(q_{\text{min}})(p_2-q_2)^4+\sigma_{11}(p_1-q_1)^2]}.
    \end{split}
\end{equation}
 Repeating a similar analysis as above gives that all contributions of the integral are irrelevant in the limit that $q_2 \rightarrow 0$. Hence $C_{1212}^R(\mathbf{q})$ becomes a constant below $\min\{q_{\sigma},q_{\text{th}}\}$.

\paragraph{Scaling Behavior of $B_{1122}$ and $B_{1111}$.}
Whereas for $C_{ijkl}^R$, we could conduct a scaling analysis corresponding to SCSA equations for the anharmonic $f^4$ interaction, $B_{1111}^R$ and $B_{1122}^R$ are coefficients of harmonic terms and will be masked by the stress in the correlation function, $\mathcal{G}_{ff}$, when $q<q_{\sigma}$.

\begin{table}
{
\begin{tabular}{ |p{2cm}|p{1cm}|p{2cm}|p{3.25cm}|  }
 \hline
 \multicolumn{4}{|c|}{Scaling Exponents $q_{\text{th}}>q_{\sigma},\sigma_{11}>0,\sigma_{22}=0$} \\
 \hline
 Scale&  $q >q_{\text{th}}$& $q_{\text{th}} > q > q_{\sigma}$& $q_{\sigma}>q$\\
 \hline
 $C_{1111}^R/C_{1111}$ &   $1$  & $\bigg( \frac{q}{q_{\text{th}}} \bigg)^{\eta_u}$ & $\bigg( \frac{q_{\sigma}}{q_{\text{th}}} \bigg)^{\eta_u} $\\
 $C_{1212}^R/C_{1212}$&     $1$ & $\bigg( \frac{q}{q_{\text{th}}} \bigg)^{\eta_u}$ & $\bigg( \frac{q_{\sigma}}{q_{\text{th}}} \bigg)^{\eta_u} $\\
 $C_{1122}^R/C_{1122}$&   $1$  & $\bigg( \frac{q}{q_{\text{th}}} \bigg)^{\eta_u}$ & $\bigg( \frac{q_{\sigma}}{q_{\text{th}}} \bigg)^{\eta_u}  \frac{q_2}{q_{\sigma}} \Omega_C^1(1/\sqrt{\tilde{q}_1})$\\
 $C_{2222}^R/C_{2222}$&   $1$  & $\bigg( \frac{q}{q_{\text{th}}} \bigg)^{\eta_u}$ & $\bigg( \frac{q_{\sigma}}{q_{\text{th}}} \bigg)^{\eta_u}  \frac{1}{\tilde{q}_{\sigma}} \Omega_C^2(1/\sqrt{\tilde{q}_1})$\\
 $B_{1111}^R/B_{1111}$ &   $1$  & $\bigg( \frac{q}{q_{\text{th}}} \bigg)^{-\eta}$ & \text{Masked}\\
 $B_{1122}^R/B_{1122}$&   $1$  & $\bigg( \frac{q}{q_{\text{th}}} \bigg)^{-\eta}$ & \text{Masked}\\
 $B_{2222}^R/B_{2222}$&   $1$  & $\bigg( \frac{q}{q_{\text{th}}} \bigg)^{-\eta}$ & $\bigg( \frac{q_{\sigma}}{q_{\text{th}}} \bigg)^{-\eta}$\\
 \hline
\end{tabular} 
}
\caption{ \label{tab:tabscaling} The scaling of the elastic moduli is shown when stress is small enough such that $q_{\text{th}} >q_{\sigma}$. When $q_{\sigma}>q$, the bending rigidities $B_{1111}^R q_1^4$ and $B_{1122}^R q_1^2 q_2^2$ are dominated by the stress term $\sigma_{11}q_1^2$ in Eq.~\ref{eq:EffectiveGreenFlexural} and we term them as masked. }
\end{table}

\begin{table}
{
\begin{tabular}{ |p{2cm}|p{1cm}|p{2cm}|p{3.25cm}|  }
 \hline
 \multicolumn{4}{|c|}{Scaling Exponents $q_{\sigma}>q_{\text{th}},\sigma_{11}>0,\sigma_{22}=0$} \\
 \hline
 Scale&  $q >q_{\sigma}$& $q_{\sigma} > q >q_{\text{th}}$& $q_{\text{th}}>q$\\
 \hline
 $C_{1111}^R/C_{1111}$ &   $1$  & $1$ & $1$\\
 $C_{1212}^R/C_{1212}$&     $1$ & $1$ & $1$\\
 $C_{1122}^R/C_{1122}$&   $1$  & $1$ & $  \frac{q_2}{q_{\text{th}}} \Omega_C^1(1/\sqrt{\tilde{q}_1})$\\
 $C_{2222}^R/C_{2222}$&   $1$  & $1$ & $ \frac{q_2}{q_{\text{th}}} \Omega_C^2(1/\sqrt{\tilde{q}_1})$\\
 $B_{1111}^R/B_{1111}$ &   $1$  & \text{Masked} & \text{Masked}\\
 $B_{1122}^R/B_{1122}$&   $1$  & \text{Masked} & \text{Masked}\\
 $B_{2222}^R/B_{2222}$&   $1$  & $1$ & $1$\\
 \hline
\end{tabular} 
}
\caption{\label{tab:tabscalinglsiglessthanlth} The scaling of the elastic moduli is shown when stress is large enough such that $q_{\text{th}} <q_{\sigma}$. When $q_{\sigma}>q$, the bending rigidities $B_{1111}^R q_1^4$ and $B_{1122}^R q_1^2 q_2^2$ are dominated by the stress term $\sigma_{11}q_1^2$  in Eq.~\ref{eq:EffectiveGreenFlexural}  and we term them as masked.} 
\end{table}

With our theoretical results we now move on to verify the scaling of this theory via simulations that measure the in-plane and flexural correlation functions in these regimes.

\begin{figure*}[t]
\includegraphics{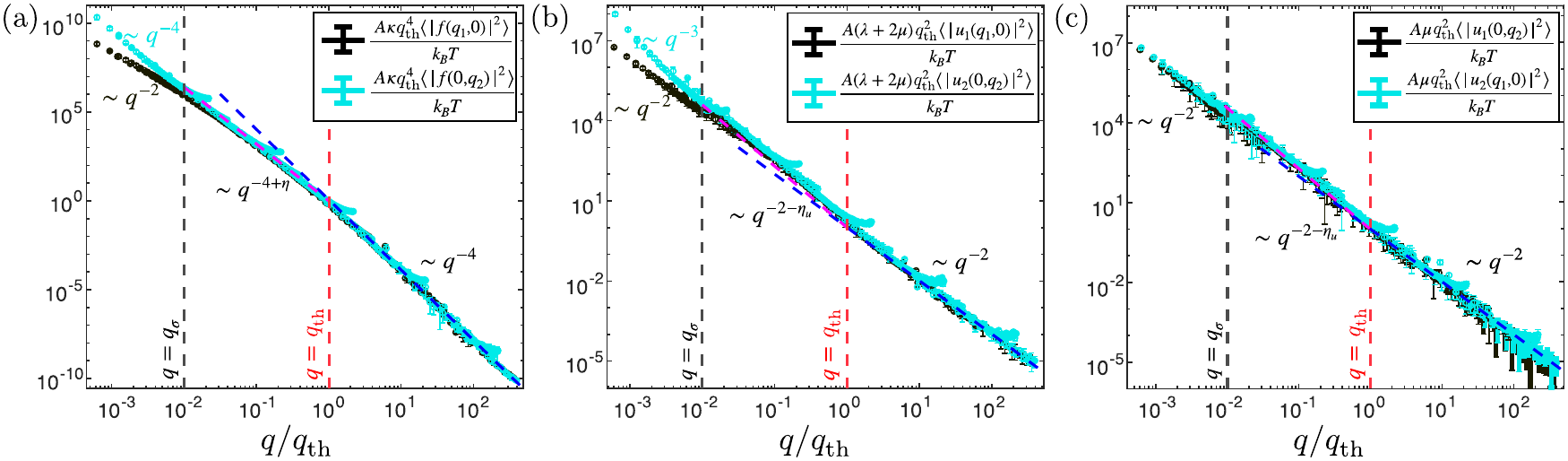}
\caption{Displayed are some simulation results for the (a) flexural correlation functions (b) orthogonal in-plane correlation functions and (c) transverse in-plane correlation function along orthogonal axes. Plots show the changes in these correlation function at the thermal transition, $q=q_{\text{th}}$, and when stress becomes dominant, $q=q_{\sigma}$. Blue dashed lines show the continuation of the harmonic scaling to aid seeing the change in slope when $q=q_{\text{th}}$. In (a) and (b), the anisotropy of the correlation functions can be observed when $q<q_{\sigma}$. The magenta lines show the anomalous thermal exponents $\eta,\eta_u$ when $q_{\text{th}}>q>q_{\sigma}$. As one can see in (c), the shear modes always remain isotropic.}
\label{fig:DoubleTransitionPanel}
\end{figure*}

\subsection{Discussion of  Correlation functions from Simulations}
The scaling of these moduli should be reflected in the correlation functions in Eqs.~\eqref{eq:EffectiveGreenFlexural} and \eqref{eq:GreenInPlanePhonon}. Molecular dynamics simulations of square-shape systems with spring-mass system arranged in a triangular lattice. Two system sizes were used, one with 2900 masses (amounting to a square sheet of size $50 a \times 50 a$ where $a$ is the lattice spacing) and 11600 masses ($100 a \times 100 a$) to show that finite size effects in the correlation functions are negligible. Further details of the simulations can be found in the appendix in Sec.~\ref{sec:MethodSimulation}. What gives the dimensional sense of system size are the parameters such as bending rigidity, Young's modulus and temperature, all of which enter into the formula for $q_{\text{th}}$. To make this clearer, in the low-stress limit:
\begin{equation}
    q_{\text{th}} \sim \sqrt{\frac{k_BTY}{\kappa^2}}
\end{equation}
where $k_BT,\kappa$ have units of energy but $Y$ has units of energy/$\text{m}^2$. Understanding this, the temperature, bending rigidity and Young's modulus were varied in order to piece together data from simulations across a large scale change, which allowed us to be more computationally effective. Specific parameters can be obtained in the appendix. Simulations were only done in the low stress limit. This is because replication of similar results in the large stress case were rendered difficult to obtain due to the non-linear responses of the lattice of springs as well.

Looking at Fig.~\ref{fig:DoubleTransitionPanel}, all simulations had a stress value such that $q_{\sigma}/q_{\text{th}} = 10^{-2}$ (using Eq.~\eqref{eq:lengthscalerelationlow}) while $\kappa,Y,T$ were varied independently. The Fig.~\ref{fig:DoubleTransitionPanel}(a) shows the transition from the harmonic regime to an anomalous thermally renormalized regime where the bending rigidities diverge isotropically with exponent $\eta \approx .8$. At $q_{\sigma}$ a second transition can be observed from the isotropic anomalous exponents $\eta,\eta_u$ to a regime where anisotropies develop and the scaling takes the form in Table~\ref{tab:tabscaling}. In-plane phonon correlation functions associated with normal strains are plotted in Fig.~\ref{fig:DoubleTransitionPanel}(b), using the same simulations. Similarly, they show the scaling expected from the theory with a strong anisotropy that develops below $q_{\sigma}$.
Finally, we also observed the isotropy of the shear modulus in Fig.~\ref{fig:DoubleTransitionPanel}(c) which also matched the scaling we found via our SCSA equations. The shear modulus ceases to renormalize once stress becomes relevant.

Having confirmed our theoretical results with simulations we can now move on to measuring the stress-strain theory that follow from this scaling theory.

\section{Simulations of Stress-Strain and Poisson's Ratio \label{sec:StressStrain}}
The stress-strain relationship of thermalized 2D sheets of dimensions $L \times L$ under uni-axial stress along axis 1 can be theoretically calculated:
\begin{equation}
    \bigg\langle \frac{\delta L_1}{L}\bigg\rangle_{\sigma} \approx \frac{\sigma_{11}}{Y} -\frac{1}{2} \sum_{\frac{2\pi}{L}<|\mathbf{q}|<\Lambda} q_1^2 \mathcal{G}_{ff}^R(\mathbf{q})
\end{equation}
where $L$ is the system size, $\Lambda$ is again the UV cutoff and $\delta L_1$ is the change in length along axis 1 \cite{kovsmrlj2016response}. The first term in the equation reflects the bare response of the material whereas the second term involves the effect of temperature. It is this latter term that gives rise to the tendency of elastic membranes to shrink \cite{kovsmrlj2016response,guitter1988crumpling,nelson2004statistical}. Similarly, strains along the axis orthogonal to the stress can be calculated as:
\begin{equation}
    \bigg\langle \frac{\delta L_2}{L}\bigg\rangle_{\sigma} \approx \frac{- \nu \sigma_{11}}{Y} -\frac{1}{2} \sum_{\frac{2\pi}{L}<|\mathbf{q}|<\Lambda} q_2^2 \mathcal{G}_{ff}^R(\mathbf{q})
\end{equation}
where $\nu$ is the bare Poisson ratio, in our case $+1/3$ for a triangular lattice. $\delta L_2$ is the change of system length along axis 2. The strains are then defined as:
\begin{equation}
\label{eq:straindifference}
    \begin{split}
        &\epsilon_{11} = \bigg\langle \frac{\delta L_1}{L}\bigg\rangle_{\sigma}-\bigg\langle \frac{\delta L_1}{L}\bigg\rangle_{0} \\
        &\epsilon_{22} = \bigg\langle \frac{\delta L_2}{L}\bigg\rangle_{\sigma}-\bigg\langle \frac{\delta L_2}{L}\bigg\rangle_{0}
    \end{split}
\end{equation}
where the subtracted terms express the reference system size in the absence of stress. These terms are necessary to subtract in order to obtain a strain from the un-stressed state where thermal fluctuations naturally induce a shrinking of the membrane.
By plugging in our theoretical scaling ansatz for the correlation functions, found in Table~\ref{tab:tabstressstrainGreen}, we can analytically calculate the stress-strain relation. Typically, for real materials such as graphene at room temperature, $a<\ell_{\text{th}}<L$ ($\ell_{\text{th}} \approx 2 \text{nm}$ at $300$K). However, since we can only effectively simulate system sizes of the order of $50a\times50a$, we tuned parameters to generally obtain a large separation of length scales $L/\ell_{\text{th}}$. We therefore examine the scaling of the stress-strain relation when $\ell_{\text{th}}<a<L$ ($2\pi/L  < \Lambda< q_{\text{th}}$) with the stress length scale being variable. The scaling ansatz of the correlation function for the length scales that fall between $a$ and $L$ depends on the magnitude of stress and is shown in Table~\ref{tab:tabstressstrainGreen}. We may show an example of how to obtain one of the scaling functions of $\epsilon_{11},\epsilon_{22}$ observed in Table~\ref{tab:tabstressstrainGreen}. 

\begin{widetext}
\begin{center}
\begin{table}[h!]
{\begin{tabular}{ |p{1cm}|p{3.2cm}|p{2.5cm}|p{3.2cm}|p{1.6cm}|  }
 \hline
 \multicolumn{5}{|c|}{Expressions of $\mathcal{G}_{ff}^R$ for $2\pi/L<\Lambda<q_{\text{th}}$} \\
 \hline Scale &
 $q_{\sigma}<2\pi/L$& $ 2\pi/L<q_{\sigma}<\Lambda $& $\Lambda<q_{\sigma}<q_{\text{th}}$&$q_{\text{th}}<q_{\sigma}$\\
 \hline
 $q_{\sigma}<q$& $\frac{k_BT}{A[\kappa q_{\text{th}}^{\eta}q^{4-\eta}+\sigma_{11}q_1^2]}$  & $\frac{k_BT}{A[\kappa q_{\text{th}}^{\eta}q^{4-\eta}+\sigma_{11}q_1^2]}$ & $\text{NA}$ & $\text{NA}$\\
 $q<q_{\sigma}$& $\text{NA}$ & $\frac{k_BT}{A[\kappa q_{\sigma}^{-\eta}q_{\text{th}}^{\eta}q_2^4+\sigma_{11}q_1^2]}$ & $\frac{k_BT}{A[\kappa q_{\sigma}^{-\eta}q_{\text{th}}^{\eta}q_2^4+\sigma_{11}q_1^2]}$ & $\frac{k_BT}{A[\kappa q_2^4+\sigma_{11}q_1^2]}$\\
 \hline
 \multicolumn{5}{|c|}{Scaling of Strains} \\
 \hline Strain &
 $ q_{\sigma}<2\pi/L$& $2\pi/L<q_{\sigma}<\Lambda $& $\Lambda<q_{\sigma}<q_{\text{th}}$&$q_{\text{th}}<q_{\sigma}$\\
 \hline
 $\epsilon_{11}$& $\sigma_{11}/(4(1-\eta)Y_R(L))$  & $\sim \sigma_{11}^{\eta/(2-\eta)}$ & $\text{transition to}$ $\sigma_{11}/Y$ & $\sigma_{11}/Y$\\
 $\epsilon_{22}$& $-\sigma_{11}/(12(1-\eta)Y_R(L))$  & $\sim \sigma_{11}^{\eta/(2-\eta)}$ & $\text{transition to}$ $-\nu \sigma_{11}/Y$ & $-\nu \sigma_{11}/Y$\\
 \hline
\end{tabular} 
}
\caption{ In this table we show the scaling of the flexural correlation functions  derived from Sec.\ref{sec:ScaleBehavior}. We then write down the corresponding stress-strain behaviors of the strains $\epsilon_{11}$ and $\epsilon_{22}$. In the table $Y_R(L) = Y(2 \pi /q_{\text{th}}L)^{\eta_u}$.}\label{tab:tabstressstrainGreen}
\end{table}
\end{center}

For example, in the case $\ell_{\text{th}}<a<\ell_{\sigma}<L (2\pi/L<q_{\sigma}<\Lambda<q_{\text{th}})$. Beginning with Eq.~\eqref{eq:straindifference}:

\begin{equation}
    \begin{split}
        &\epsilon_{11}(\sigma_{11},T | 2\pi/L<q_{\sigma}<\Lambda<q_{\text{th}}) \\= & \frac{\sigma_{11}}{Y} - \frac{1}{2} \sum_{q_{\sigma} < |\mathbf{q}|< \Lambda  } q_1^2 \bigg[ \frac{k_BT}{A[\kappa q_{\text{th}}^{\eta} q^{4-\eta} +\sigma_{11} q_1^2 ]} - \frac{k_BT}{A\kappa q_{\text{th}}^{\eta} q^{4-\eta} } \bigg]  - \frac{1}{2} \sum_{\frac{2 \pi}{L} < |\mathbf{q}|< q_{\sigma} } q_1^2 \bigg[ \frac{k_BT}{A[\kappa q_{\sigma}^{-\eta}q_{\text{th}}^{\eta} q_2^{4} +\sigma_{11} q_1^2 ]}  - \frac{k_BT}{A\kappa q_{\text{th}}^{\eta} q^{4-\eta} }\bigg] \\ = & 
        \frac{\sigma_{11}}{Y} - \frac{1}{2} \sum_{q_{\sigma} < |\mathbf{q}|< \Lambda  } q_1^2 \bigg[ - \frac{k_BT \sigma_{11} q_1^2}{A\kappa^2 q_{\text{th}}^{2\eta} q^{8-2\eta}} \bigg] - \frac{1}{2} \sum_{\frac{2 \pi}{L} < |\mathbf{q}|< q_{\sigma} } q_1^2 \bigg[ \frac{k_BT}{A[\kappa q_{\sigma}^{-\eta}q_{\text{th}}^{\eta} q_2^{4} +\sigma_{11} q_1^2 ]}  - \frac{k_BT}{A\kappa q_{\text{th}}^{\eta} q^{4-\eta} }\bigg]
    \end{split}
\end{equation}
and in the first summation we may Taylor expand the correlation function to first order (since for those wave vectors the stress term is not dominant in the denominator of the flexural correlation function). We can then convert these terms to integrals:
\begin{equation}
    \begin{split}
         \epsilon_{11}(\sigma_{11},T | 2\pi/L<q_{\sigma}<\Lambda<q_{\text{th}}) \approx &\frac{\sigma}{Y} - \frac{1}{2(2\pi)^2} \int_{0}^{2\pi} d \theta \int_{q_{\sigma}}^{\Lambda}dq q^3 \cos^2{\theta} \bigg[ - \frac{k_BT \sigma_{11} q^2 \cos^2{\theta}}{\kappa^2 q_{\text{th}}^{2\eta} q^{8-2\eta}} \bigg]
         \\ &- \frac{1}{2(2\pi)^2} \int_{0}^{2\pi} d \theta \int_{2\pi/L}^{q_{\sigma}}dq q^3 \cos^2{\theta}  \bigg[ \frac{k_BT}{[\kappa q_{\sigma}^{-\eta}q_{\text{th}}^{\eta} q^{4}\sin^4{\theta} +\sigma_{11} q^2 \cos^2{\theta} ]} - \frac{k_BT}{\kappa q_{\text{th}}^{\eta} q^{4-\eta} }\bigg]
    \end{split}
\end{equation}
To make the latter integral tractable we approximate the denominator in the following manner:
\begin{equation}
    \begin{split}
    \frac{k_BT}{A[\kappa q_{\sigma}^{-\eta}q_{\text{th}}^{\eta} q^{4}\sin^4{\theta} +\sigma_{11} q^2 \cos^2{\theta} ]} \approx \frac{k_BT}{A[\kappa q_{\sigma}^{-\eta}q_{\text{th}}^{\eta} q^{4} +\sigma_{11} q^2 \cos^2{\theta} ]}
    \end{split}
\end{equation}
\end{widetext}
This approximation is justified since the stress is dominant when $\theta \neq \pi/2$ in the domain of the integral, $q \in [2\pi/L,q_{\sigma}]$. These integrals can now be analytically integrated giving rise to:
\begin{widetext}
\begin{equation}
 \begin{split}
    \epsilon_{11}\left(\sigma_{11},T \bigg| \frac{2\pi}{L}<q_{\sigma}<\Lambda<q_{\text{th}}\right) = \frac{\sigma_{11}}{Y}  & -\bigg[ \frac{3k_BT \sigma_{11}}{64 \pi (1-\eta)\kappa^2q_{\text{th}}^{2\eta}} q^{2 \eta -2}\bigg] \bigg|_{q_{\sigma}}^{\Lambda} \\ &- \frac{k_BT}{8 \pi \sigma_{11}} \bigg[q^2 - q \sqrt{q^2+ \bigg(\frac{q_{\sigma}}{q_{\text{th}}}\bigg)^{\eta} \frac{\sigma_{11}}{\kappa}} + \bigg( \frac{q_{\sigma}}{q_{\text{th}}}\bigg)^{\eta} \frac{\sigma_{11}}{\kappa} \text{sinh}^{-1}\bigg[q\bigg( \frac{q_{\sigma}}{q_{\text{th}}}\bigg)^{-\eta/2} \sqrt{\frac{\kappa}{\sigma_{11}}} \bigg] \bigg] \bigg|_{\frac{2 \pi}{L}}^{q_{\sigma}}\\
    &+ \frac{k_BT}{8 \pi \eta \kappa} \bigg(\frac{q}{q_{\text{th}}}\bigg)^{\eta} \bigg|_{\frac{2\pi}{L}}^{q_{\sigma}}
 \end{split} 
\end{equation}
\end{widetext}
By taking the infinite system size limit (which is appropriate since we are also assuming large separation of length scales that $\frac{2 \pi}{L} \bigg( \frac{q_{\sigma}}{q_{\text{th}}}\bigg)^{-\eta/2} \sqrt{\frac{\kappa}{\sigma}}  \ll 1 $ when $2\pi/L<q_{\sigma}<\Lambda<q_{\text{th}}$), we can obtain a simpler expression:

\begin{widetext}
\begin{equation} \label{eq:epsilon11}
 \begin{split}
    \lim_{L \rightarrow \infty}\epsilon_{11}\bigg(\sigma_{11},T \bigg| &\frac{2\pi}{L}<q_{\sigma}<  \Lambda<q_{\text{th}}\bigg)  \\ &= \frac{\sigma_{11}}{Y}\bigg[1 -   \frac{1}{2 (1-\eta)} \bigg(\frac{\Lambda}{q_{\text{th}}} \bigg)^{2 \eta -2}\bigg]  - \frac{k_BT}{8 \pi \kappa}\bigg( \frac{q_{\sigma}}{q_{\text{th}}}\bigg)^{\eta} \bigg[(1-\sqrt{2}) + \text{sinh}^{-1}(1) -\eta^{-1} - \frac{3}{8(1-\eta)} \bigg]  \\
    & = \frac{\sigma_{11}}{Y}\bigg[1 -   \frac{1}{2 (1-\eta)} \bigg(\frac{\Lambda}{q_{\text{th}}} \bigg)^{2 \eta -2}\bigg]  - \frac{k_BT}{8 \pi \kappa}\bigg( \frac{16 \pi \sigma_{11} \kappa}{3k_BTY}\bigg)^{\frac{\eta}{2-\eta}} \bigg[(1-\sqrt{2}) + \text{sinh}^{-1}(1) -\eta^{-1} - \frac{3}{8(1-\eta)} \bigg]
 \end{split} 
\end{equation}
A similar calculation gives: 
\begin{equation} \label{eq:epsilon22}
 \begin{split}
    \lim_{L \rightarrow \infty}\epsilon_{22}\bigg(\sigma_{11},T \bigg| &\frac{2\pi}{L}<q_{\sigma}<  \Lambda<q_{\text{th}}\bigg)  \\
    & = \frac{ \sigma_{11}}{Y}\bigg[-\nu -   \frac{1}{6 (1-\eta)} \bigg(\frac{\Lambda}{q_{\text{th}}} \bigg)^{2 \eta -2}\bigg]  - \frac{k_BT}{8 \pi \kappa}\bigg( \frac{16 \pi \sigma_{11} \kappa}{3k_BTY}\bigg)^{\frac{\eta}{2-\eta}} \bigg[(-1+\sqrt{2}) + \text{sinh}^{-1}(1) -\eta^{-1} - \frac{1}{8(1-\eta)} \bigg]
 \end{split} 
\end{equation}
\end{widetext}

The rest of the strains for other regimes can be obtained in a similar manner and are shown in Table~\ref{tab:tabstressstrainGreen} with explicit solutions in Sec.~\ref{sec:StressStrainApp}, and these scalings become more accurate with a large separation of length scales (in other words if $2\pi/L$, $q_{\text{th}}$ and $q_{\sigma}$ being all different orders of magnitude) \cite{burmistrov2018stress}. 

Within Eqs.~\eqref{eq:epsilon11} and \eqref{eq:epsilon22}, pre-factors of each of the power laws can be compared using the fact that $\sigma_{q_{\text{th}}} \gg \sigma$ (where $\sigma_{q_{\text{th}}}$ is defined as the stress such that $q_{\sigma} = q_{\text{th}}$). The comparison shows that the last terms in each equation, which exhibit the scaling $\sigma^{\eta/(2-\eta)}$, is the dominant power law.
Thus, when $2\pi/L<q_{\sigma}<\Lambda$ with $L \rightarrow \infty$ and holding stress fixed, a non-linear stress-strain regime appears for both $\epsilon_{11}$ and $\epsilon_{22}$. This scaling for the strains was already known in Ref. \cite{kovsmrlj2016response, guitter1989thermodynamical}. Therefore in the same stress regime, we expect to have a universal absolute Poisson ratio value since:
\begin{equation}
\nu^R = -\frac{\epsilon_{22}}{\epsilon_{11}} \approx -\frac{-1+\sqrt{2}+\arcsinh^{-1}(1)-\eta^{-1}-\frac{1}{8(1-\eta)}}{1-\sqrt{2}+\arcsinh^{-1}(1)-\eta^{-1}-\frac{3}{8(1-\eta)}}   
\end{equation}
Theoretically this is expected when the separation of length scales is sufficiently large \cite{burmistrov2018stress}. Like \cite{burmistrov2018stress,saykin2020absolute}, our value, plugging in $\eta \approx .8$, would not match with the linear response value of $-1/3$. Previous theoretical investigations that have obtained this linear response of $-1/3$, have calculated it via the elastic moduli $\lambda^R/(\lambda^R+2\mu^R)$ which is governed by the Aronovitz-Lubensky fixed point \cite{le1992self,gazit2009structure,hasselmann2011nonlocal}.  \cite{burmistrov2018stress,saykin2020absolute,burmistrov2018differential} find that the differential Poission ratio is $-1/3$ in the non-linear regime only when $d_c \rightarrow \infty$ whereas the absolute Poisson ratio is never $-1/3$. In addition, the Poisson ratio is sensitive to the type of boundary condition that is used \cite{burmistrov2018stress}.

\begin{figure}[h!]
\includegraphics{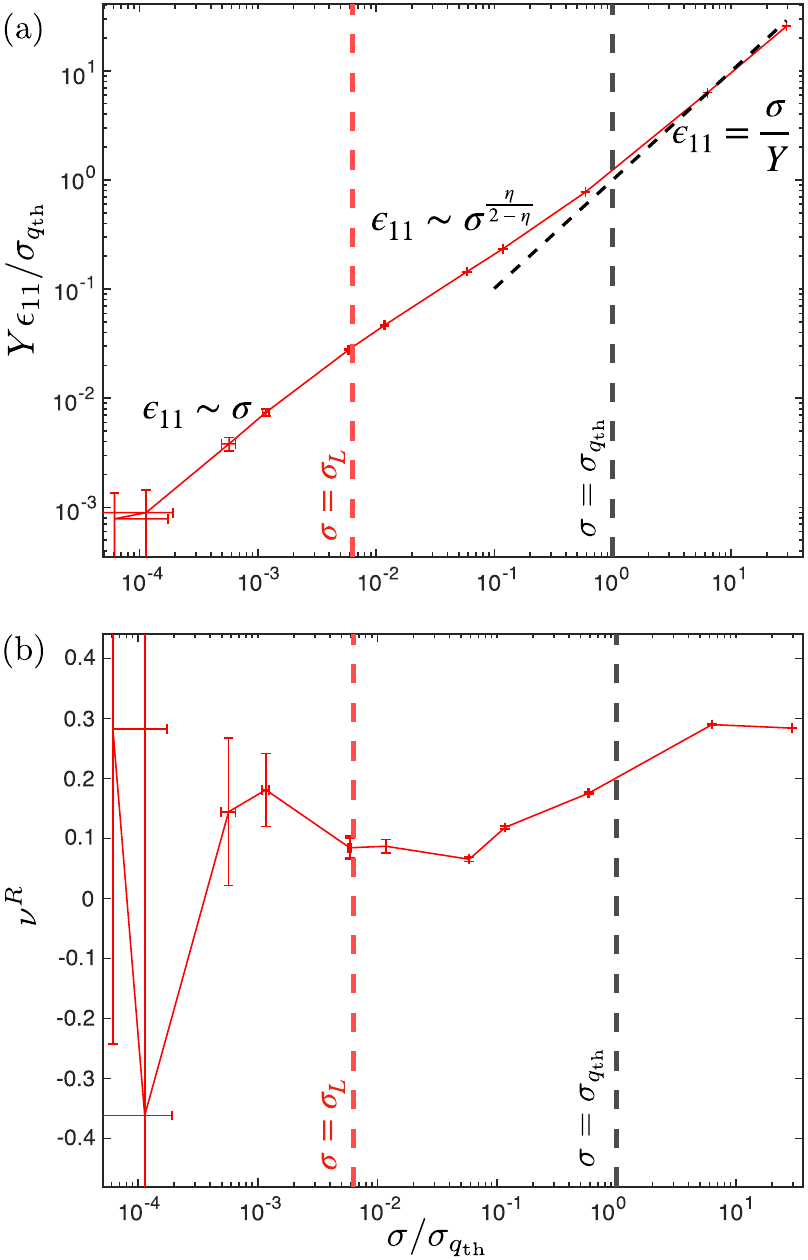}
\caption{ The stress strain curve (a) and Poisson strain curve (b) are plotted for simulations with system size $50a\times50a$, only changing the value of stress. $\sigma_{q_{\text{th}}}$ is defined as the stress at which $q_{\sigma} = q_{\text{th}}$. The red dashed vertical line marks when $\sigma = \sigma_L$ (which is when $q_{\sigma}=2\pi/L$, a non-linear regime where $\epsilon_{11} \sim \sigma^{.72}$ appears. The angled dashed line marks the $y=x$ line and shows that for large stresses, a classical response is regained.}
\label{fig:StressPoissonStrainPlot}
\end{figure}

With simulations we first sought to confirm the non-linear stress strain relation, which can be observed in Fig.~\ref{fig:StressPoissonStrainPlot} (a). Between $\sigma_L$ and $\sigma_{q_\text{th}}$ we observed this non-linear relation. For large stresses, the classical response absent of any effects of thermal fluctuations (when $\sigma>\sigma_{q_{\text{th}}}$ in other words when $q_{\text{th}}<q_{\sigma}$) is obtained. For very small stresses, and with a large separation of length scales, one should observe a linear response that follows from $\epsilon_{11} \approx \sigma/4(1-\eta)Y_R(L)$, where $Y_R(L) = Y(2 \pi /q_{\text{th}}L)^{\eta_u}$ (explained in Sec.~\ref{sec:StressStrainApp}). We were not able to numerically verify this slope, however we do observe a linear theory where the Young's modulus is softened by thermal fluctuations.

From the same simulations, we can obtain the Poisson ratio by taking the negative ratio of strains $\epsilon_{11}$ and $\epsilon_{22}$. In Fig.~\ref{fig:StressPoissonStrainPlot}(b), the Poisson ratio is plotted against the stress. The Poisson ratio shows potentially a universal flat regime for stress values such that $2\pi/L<q_{\sigma}<q_{\text{th}}$ \cite{burmistrov2018stress}. For very small stresses, errors became very difficult to control. Stress free Monte-Carlo simulations in the past, \cite{falcioni1997poisson}, did measure a Poisson ratio via correlations functions and found a linear response of $-1/3$ predicted by \cite{le1992self}. Evidence from other simulations is much more scattered however. In \cite{zhang1996molecular}, the Poisson ratio was measured to be $-.15$. More recent simulations in \cite{fasolino2007intrinsic} may show that the linear response Poisson ratio may be positive. Further simulations done by \cite{los2016scaling} also found a disagreement with the value of the $-1/3$ in the thermodynamic limit. Thus it is unclear as to what should be the precise value of both the linear response of the Poisson ratio as well as its behavior in the non-linear regime.

Returning to our own data, for large stresses such that $\sigma>\sigma_{q_{\text{th}}}$, the bare Poisson ratio of the triangular lattice of masses connected by springs, $1/3$, could not be achieved due to the immediate cross-over to the non-linear elastic regime (in the simulations the data showing the box length along the axis of stress begins to become very large at these stresses leading to a decrease in the Poisson ratio with further application of stress).

\section{Conclusions}
We examined the effects of uni-axial stress on thermally fluctuating sheets. In particular, we see that anomalous scaling due to thermal fluctuations at scales where the uni-axial stress is dominant still appears in the in-plane moduli orthogonal to the stress, such as $C_{2222}^R$. Furthermore the presence of the two length scales $q_{\sigma}$ and $q_{\text{th}}$ provides an interesting foreground for various regimes of the scaling of moduli. We verified these scalings via simulations, in particular the anomalous scaling of $C_{2222}^R$ as well as the transition at $q_{\sigma}$, beyond which the correlation functions becomes anisotropic. These results match with previous investigations of tubules \cite{radzihovsky1995new}.

We furthermore verified the existence of a non-linear stress-strain regime $\epsilon \sim \sigma^{\eta/(2-\eta)}$ in our simulations with a numerically accurate exponent. However our results measuring the Poisson ratio were less conclusive and require further investigation.

\appendix
\section{Supplementary Information}

\subsection{Methods of Simulation \label{sec:MethodSimulation}}
Simulations were performed on a cluster using 2.4 GHz Broadwell CPUs using molecular dynamics package LAMMPS in the NPT ensemble using a Nos\'e-Hoover thermostat. The simulations were of a 2D isotropic spring-mass triangular lattice embedded in 3 dimensions and under periodic boundary conditions. The elastic bending energy of such a spring mass system can be formulated as:
\begin{equation}
    \text{E}_{\text{bend}} = \frac{\hat{\kappa}}{2} \sum_{\langle IJ \rangle } [1+\text{cos}\theta_{IJ}]
\end{equation}
where $\hat{\kappa}$ is the microscopic dihedral spring stiffness and $\theta_{IJ}$ is the dihedral angle between two triangular faces (which can also be seen as the angle differences between normals of faces). The stretching energy is instead:
\begin{equation}
    \text{E}_{\text{stretch}} = \frac{\hat{Y}}{2} \sum_{\langle ij \rangle } (r_{ij}-a)^2
\end{equation}
where $r_{ij} = |\mathbf{r}_i - \mathbf{r}_j|$ is the Euclidean distance between two neighbors $i$ and $j$ and $a$ is the lattice spacing. The bare continuum moduli of such a system can be derived from the discrete spring stiffnesses \cite{seung1988defects}:
\begin{equation}
    \kappa = \frac{\sqrt{3}}{2} \hat{\kappa} , \lambda = \mu = \frac{\sqrt{3}}{4} \hat{Y}
\end{equation}
The parameters were generally varied and hence the time step had to be chosen carefully to be less or equal to the following reduced times and periods:
\begin{equation}
    \tau_T = a \sqrt{\frac{m}{k_BT}}, \tau_{\hat{Y}} = \sqrt{\frac{m}{\hat{Y}}}, \tau_{\hat{\kappa}} = a \sqrt{\frac{m}{\hat{\kappa}}}
\end{equation}
The simulations were done non-dimensionally so $k_B$, the Boltzman constant, and mass and lattice spacing were set to $1$. A simulation generally ran for approximately $1.6 \times 10^8 - 10^9$ time steps each of length $\text{Min} \{ \tau_T, \tau_{\hat{Y}} , \tau_{\hat{\kappa}} \}$. In computation time this equates to 6-60 hours on the cluster. The system size was mostly kept constant around $50 \times 50$ and $100 \times 100$ for the molecular dynamics correlations. 

\subsection{Data Sets}

\begin{table}[h!]
{\caption{\label{tab:tabdatafig2}}
\begin{tabular}{ |p{1cm}|p{1cm}|p{2cm}|p{2cm}|  }
 \hline
 \multicolumn{4}{|c|}{Data Sets for Fig.~\ref{fig:DoubleTransitionPanel},$q_{\sigma}/q_{\text{th}} = 10^{-2}$} \\
 \hline
 $L/a$ &  $\hat{\kappa}/k_BT$& $\hat{Y}/(k_BT/a^2)$& $\hat{\sigma}/(k_BT/a^2)$\\
 \hline
 $50 $ &   $10^3 $  & $220$ & $1.4 \times 10^{-4}$ \\
 $50 $&     $10^2$ & $220$ & $1.4 \times 10^{-3}$ \\
 $50 $&   $10^2$  & $2.2 \times 10^4$ & $.14$ \\
 $50$&   $1$  & $220$& $.14$ \\
 $50$ &   $1$  & $2.2 \times 10^4$& $14$ \\
 $50$&   $1$  & $2.2 \times 10^5$& $140$ \\
 $100$&   $10^3$  & $2.2 \times 10^3$& $5.6\times 10^{-3}$ \\
 $100$&   $10^2$  & $2.2 \times 10^3$&$5.6\times 10^{-2}$ \\
 $100$&   $10^2$  & $2.2 \times 10^5$ &$5.6$\\
 $100$&   $1$  & $2.2 \times 10^3$&$5.6$ \\
 $100$&   $1$  & $2.2 \times 10^5$ &$560$\\
 \hline
\end{tabular} 
}
\end{table}

\subsection{Homogeneous Integrals For SCSA Analysis of \texorpdfstring{$C_{1111}^R,C_{1122}^R$}{TEXT} \label{sec:Integrals}}

\begin{widetext}
\begin{equation}
    \begin{split}
     & I_{1122}^{(1)} = -\frac{k_BT}{4(2\pi)^2} \int d \tilde{p}_1 d \tilde{p}_2  \frac{(\tilde{p}_2-1)\tilde{p}_2 (\tilde{p}_1-\tilde{q}_1)\tilde{p}_1}{[B_{2222}^R(q_{\text{min}})(\tilde{p}_2-1)^4+\sigma_{11}(\tilde{p}_1-\tilde{q}_1)^2][B_{2222}^R(q_{\text{min}})\tilde{p}_2^4+\sigma_{11}\tilde{p}_1^2]}
     \\ &
     I_{1122}^{(2)} = -\frac{2k_BT}{4(2\pi)^2} \int d \tilde{p}_1 d \tilde{p}_2  \frac{C_{2222}(q_{\min})\Omega_C^2(\frac{1}{\sqrt{\tilde{q}_1}})}{q_{\min}}\frac{(\tilde{p}_2-1)^2\tilde{p}_2 ^2}{[B_{2222}^R(q_{\text{min}})(\tilde{p}_2-1)^4+\sigma_{11}(\tilde{p}_1-\tilde{q}_1)^2][B_{2222}^R(q_{\text{min}})\tilde{p}_2^4+\sigma_{11}\tilde{p}_1^2]}
      \\ &
     I_{1122}^{(3)} = -\frac{k_BT}{4(2\pi)^2} \int d \tilde{p}_1 d \tilde{p}_2  \frac{C_{2222}}{C_{1122}}\frac{(\tilde{p}_2-1)^2\tilde{p}_2 ^2}{[B_{2222}^R(q_{\text{min}})(\tilde{p}_2-1)^4+\sigma_{11}(\tilde{p}_1-\tilde{q}_1)^2][B_{2222}^R(q_{\text{min}})\tilde{p}_2^4+\sigma_{11}\tilde{p}_1^2]}
     \\ &
     I_{1111}^{(1)} = -\frac{k_BT}{4(2\pi)^2} \int d \tilde{p}_1 d \tilde{p}_2  \frac{C_{1122}}{C_{1111}}\frac{(\tilde{p}_2-1)^2\tilde{p}_2 ^2}{[B_{2222}^R(q_{\text{min}})(\tilde{p}_2-1)^4+\sigma_{11}(\tilde{p}_1-\tilde{q}_1)^2][B_{2222}^R(q_{\text{min}})\tilde{p}_2^4+\sigma_{11}\tilde{p}_1^2]}
     \\ &
     I_{1111}^{(2)} = -\frac{k_BT}{4(2\pi)^2} \int d \tilde{p}_1 d \tilde{p}_2  \frac{(\tilde{p}_2-1)\tilde{p}_2 (\tilde{p}_1-\tilde{q}_1)\tilde{p}_1}{[B_{2222}^R(q_{\text{min}})(\tilde{p}_2-1)^4+\sigma_{11}(\tilde{p}_1-\tilde{q}_1)^2][B_{2222}^R(q_{\text{min}})\tilde{p}_2^4+\sigma_{11}\tilde{p}_1^2]}
     \\ &
     I_{1111}^{(3)} = -\frac{k_BT}{4(2\pi)^2} \int d \tilde{p}_1 d \tilde{p}_2  \frac{C_{1122}}{C_{1111}}\frac{(\tilde{p}_2-1)\tilde{p}_2 (\tilde{p}_1-\tilde{q}_1)\tilde{p}_1}{[B_{2222}^R(q_{\text{min}})(\tilde{p}_2-1)^4+\sigma_{11}(\tilde{p}_1-\tilde{q}_1)^2][B_{2222}^R(q_{\text{min}})\tilde{p}_2^4+\sigma_{11}\tilde{p}_1^2]}
     \\ &
     I_{1111}^{(4)} = -\frac{k_BT}{4(2\pi)^2} \int d \tilde{p}_1 d \tilde{p}_2  \frac{C_{1122}}{C_{1111}}\frac{ (\tilde{p}_1-\tilde{q}_1)^2\tilde{p}_1^2}{[B_{2222}^R(q_{\text{min}})(\tilde{p}_2-1)^4+\sigma_{11}(\tilde{p}_1-\tilde{q}_1)^2][B_{2222}^R(q_{\text{min}})\tilde{p}_2^4+\sigma_{11}\tilde{p}_1^2]}
    \end{split}
\end{equation}
\end{widetext}

\subsection{Stress Strain Relations \label{sec:StressStrainApp}}
In this section we summarize the results of how the strains scale with respect to an applied stress for the other two relevant regimes of stress. For small stresses, when $q_{\sigma}<\frac{2\pi}{L}<\Lambda<q_{\text{th}}$ we obtain:
\begin{widetext}
\begin{equation}
 \begin{split}
    \epsilon_{11}(\sigma_{11},T | q_{\sigma}<2\pi/L<\Lambda<q_{\text{th}}) & \approx  \frac{\sigma_{11}}{Y} - \frac{1}{2} \sum_{2\pi/L < |\mathbf{q}|< \Lambda  } q_1^2 \bigg[ \frac{k_BT}{A[\kappa q_{\text{th}}^{\eta} q^{4-\eta} +\sigma_{11} q_1^2 ]} - \frac{k_BT}{A[\kappa q_{\text{th}}^{\eta} q^{4-\eta} ]}\bigg] \\
    &\approx \frac{\sigma_{11}}{Y} - \frac{1}{2(2\pi)^2} \int_{0}^{2\pi} d \theta \int_{2\pi/L}^{\Lambda}dq q^3 \cos^2{\theta} \bigg[ - \frac{k_BT \sigma_{11} q^2 \cos^2{\theta}}{\kappa^2 q_{\text{th}}^{2\eta} q^{8-2\eta}} \bigg] \\
    &= \frac{\sigma_{11}}{Y}   -\bigg[ \frac{3k_BT \sigma_{11}}{64 \pi (1-\eta)\kappa^2q_{\text{th}}^{2\eta}} q^{2 \eta -2}\bigg] \bigg|_{2\pi/L}^{\Lambda} \approx \frac{3k_BT \sigma_{11}}{64 \pi (1-\eta)\kappa^2q_{\text{th}}^{2\eta}} \bigg( \frac{2\pi}{L}\bigg)^{2 \eta -2}\\
    &\approx \frac{\sigma_{11}}{4(1-\eta)Y_R(L)}\\
    \\
        \epsilon_{22}\left(\sigma_{11},T \bigg| q_{\sigma}<\frac{2\pi}{L}<\Lambda<q_{\text{th}}\right) & \approx \frac{-\nu \sigma_{11}}{Y}   -\bigg[ \frac{k_BT \sigma_{11}}{64 \pi (1-\eta)\kappa^2q_{\text{th}}^{2\eta}} q^{2 \eta -2}\bigg] \bigg|_{2\pi/L}^{\Lambda} \approx \frac{k_BT \sigma_{11}}{64 \pi (1-\eta)\kappa^2q_{\text{th}}^{2\eta}} \bigg( \frac{2\pi}{L}\bigg)^{2 \eta -2}\\
        &\approx \frac{\sigma_{11}}{12(1-\eta)Y_R(L)}
 \end{split} 
\end{equation}
At low stresses we can ignore the bare response term $\sigma/Y$ for $\epsilon_{11}$ or $-\nu \sigma/Y$ for $\epsilon_{22}$  and since we are not interested in the effects of microscopic physics, the dominant term in the above expressions is the one that involves the system size. This term can then be reformulated in terms of the renormalized Young's modulus, $Y_R(L) = Y(2 \pi /q_{\text{th}}L)^{\eta_u}$. In addition, one can immediately see from the definition of the Poisson ratio, $\nu^R = -1/3$ in the linear response. We do not see this linear response value in our simulations however and there is theory that supports other values \cite{burmistrov2018stress}. Instead, at large stresses when $\frac{2\pi}{L}<\Lambda<q_{\text{th}}<q_{\sigma}$ we obtain:
\begin{equation}
 \begin{split}
    \epsilon_{11}(\sigma_{11},T | 2\pi/L<\Lambda<q_{\text{th}}<q_{\sigma}) & \approx  \frac{\sigma_{11}}{Y} - \frac{1}{2} \sum_{2\pi/L < |\mathbf{q}|< \Lambda  } q_1^2 \bigg[ \frac{k_BT}{A[\kappa  q^{4} +\sigma_{11} q_1^2 ]} - \frac{k_BT}{A[\kappa q_{\text{th}}^{\eta} q^{4-\eta} ]}\bigg] \\
    &=\frac{\sigma_{11}}{Y} - \frac{1}{2(2\pi)^2} \int_{0}^{2\pi} d \theta \int_{2\pi/L}^{\Lambda}dq q^3 \cos^2{\theta} \bigg[ \frac{k_BT}  {\kappa q^4 +\sigma_{11}q^2  \cos^2{\theta}} - \frac{k_BT}{\kappa q_{\text{th}}^{\eta} q^{4-\eta} }\bigg] \\
    &= \frac{\sigma_{11}}{Y}  - \frac{k_BT}{8 \pi \sigma_{11}} \bigg[q^2 - q \sqrt{q^2+  \frac{\sigma_{11}}{\kappa}} +  \frac{\sigma_{11}}{\kappa} \text{sinh}^{-1}\bigg[q \sqrt{\frac{\kappa}{\sigma_{11}}} \bigg] \bigg] \bigg|_{\frac{2 \pi}{L}}^{\Lambda}+ \frac{k_BT}{8 \pi \eta \kappa} \bigg(\frac{q}{q_{\text{th}}}\bigg)^{\eta} \bigg|_{\frac{2\pi}{L}}^{\Lambda} \\
    &\approx \frac{\sigma_{11}}{Y}
    \\
        \epsilon_{22}\left(\sigma_{11},T \bigg| \frac{2\pi}{L}<\Lambda<q_{\text{th}}<q_{\sigma}\right) & \approx  \frac{-\nu \sigma_{11}}{Y}     - \frac{k_BT}{8 \pi \sigma_{11}} \bigg[-q^2 + q \sqrt{q^2+  \frac{\sigma_{11}}{\kappa}} +  \frac{\sigma_{11}}{\kappa} \text{sinh}^{-1}\bigg[q \sqrt{\frac{\kappa}{\sigma_{11}}} \bigg] \bigg] \bigg|_{\frac{2 \pi}{L}}^{\Lambda}+ \frac{k_BT}{8 \pi \eta \kappa} \bigg(\frac{q}{q_{\text{th}}}\bigg)^{\eta} \bigg|_{\frac{2\pi}{L}}^{\Lambda} \\&\approx \frac{-\nu \sigma_{11}}{Y}
 \end{split} 
\end{equation}
\end{widetext}
where we have approximated that when stress is quite high, the terms from the integrals can be ignored and the bare material properties can be used to obtain the effective mechanical response. Thus the Poisson ratio we should observe should also be that of the bare material. For our simulations with triangular lattices, $\nu = 1/3$.
\bibliographystyle{ieeetr}
\bibliography{./refer.bib}

\end{document}